\documentclass[onecolumn]{aastex7}

\shorttitle{MHD simulations of flares}
\shortauthors{Rempel et al.}

\renewcommand{\vec}[1]{\mathbf{#1}}

\usepackage{amsmath}
\DeclareMathOperator{\arcsinh}{arcsinh}
\DeclareMathOperator{\erf}{erf}

\begin{document}

\title{Data-inspired simulation of AR 11158}

\author[0000-0001-5850-3119]{Matthias Rempel}
\affiliation{High Altitude Observatory, NSF National Center for Atmospheric Research, P.O. Box 3000, Boulder, CO 80307, USA}
\email{rempel@ucar.edu}

\author[0000-0002-1253-8882]{Georgios Chintzoglou}
\affiliation{Lockheed Martin Solar and Astrophysics Laboratory, 3251 Hanover Street, Palo Alto, CA 94304, USA}
\email{gchintzo@lmsal.com}


\begin{abstract}
We present a data-inspired simulation of NOAA active region AR 11158. We simulate the formation of a collisional polarity inversion line (cPIL) by moving sunspots in a quadrupolar configuration along the centroid positions extracted from AR 11158. This process builds up free energy in the corona exceeding $4\cdot 10^{32}$ erg, out of which about $2\cdot 10^{32}$ erg are released in a X-flare followed by a series of smaller flares in the B to M range. The 4 strongest flares are associated with coronal mass ejections. About 1-2 hours prior to the X-flare a magnetic flux rope (MFR) is forming above the cPIL. About 5 minutes before the flare an upflow at chromospheric heights indicates a rise of the MFR. The eruption starts when parts of the MFR enter regions with a decay index larger than 1.5, indicating that the flare initiation is consistent with the torus instability. Comparing the series of flares in this simulation to properties of observed flares, we find a comparable trend between flare energy and GOES X-ray flux, but flare duration falls into the short end of the observed solar distribution. As a consequence, energy fluxes into the flare ribbons can be substantial, reaching $10^{13}$ erg cm$^{-2}$ s$^{-1}$ for the simulated X-flare. The X-flare causes step-function like changes of the horizontal magnetic field in the photosphere, a propagation of a momentum pulse into the convection zone and quasi-periodic pulsations in the volume of the corona with periods from sub-seconds to multiple 10 seconds.
\end{abstract}


\section{Introduction} \label{sec:intro}
Active regions (ARs) with a complex topology are statistically more likely to produce strong flares \citep{Schrijver:2007}. AR complexity is typically characterized by the so-called Mt. Wilson (or Hale) classification, describing the complexity of the sunspot groups as seen in white-light (e.g., $\alpha$: single/dominant polarity region; $\beta$: bipolar region; $\gamma$: region of intermixed polarities). Oftentimes, the complexity is further quantified by the distribution of properties of the polarity inversion lines (PILs), such as their length, field gradients, shear angle of the field etc., which do provide some statistical likelihood for flare productivity \citep[e.g.][]{Gallagher:etal:2002,Barnes:etal:2007,Georgoulis:Rust:2007}. Through which physical pathways AR complexity translates into flare productivity remains an open question. Collisional shearing was introduced by \citet{Chintzoglou:etal:2019} as a possible scenario for complex ARs that host multiple flux emergence events of bipolar pairs that subsequently interact with each other and cause sunspots with opposite polarities to collide and shear off against each other. During that process sunspots with opposite polarities can get close enough to be classified as a $\delta$-sunspot \citep{Kuenzel:1960}. Several models have addressed how sub-photospheric dynamics can lead to the formation of $\delta$-sunspots through the emergence of twisted kink-unstable flux ropes \citet{Linton:etal:1998,Linton:etal:1999,Takasao:etal:2015} or the interaction of magnetic flux with downflows resulting from mass drainage during flux emergence \citet{Toriumi:etal:2014,Fang:Fan:2015}. A more systematic study of the various scenarios that can lead to $\delta$-spots was conducted by \citet{Toriumi:Takasao:2017}. All these investigations focused on the photospheric and sub-photospheric evolution, how they impact the corona and play a role in energizing specifically strong flares was studied by \citet{Chintzoglou:etal:2019}, who introduced the general concept of collisional shearing to capture $\delta$-spot evolution. Collisional shearing refers to situations where multiple AR-forming flux bundles emerge in close proximity and interact with each other. Of special interest here are interactions at the PILs that separate the polarities belonging to different flux bundles, also referred to as the nonconjugated polarities (with polarities belonging to bipolar flux bundles referred to as conjugated polarities). These PILs are referred to as \emph{collisional Polarity Inversion Lines} (cPILs). A recent observational analysis of 28 $\delta$-sunspots by \citet{Moore:etal:2026ApJ:delta} revealed that the majority of them (27) resulted from the interaction of at least 2 emerging bipolar active regions. \citet{Chintzoglou:etal:2019} used magneto-frictional modeling to demonstrate that collisional shearing does lead to the formation of pre-eruption magnetic flux ropes (MFR) and recurrent flaring above the cPILs strongly suggesting that collisional shearing is a potent mechanism leading to major solar eruptions. 

A full radiative MHD treatment that included the full extent from sub-photospheric to coronal layers was presented by \citet{Rempel:etal:2023}. They considered a simplified bipolar setup with imposed sub-photospheric footpoint motions that led to collisional shearing. While this setup allowed to study the interaction of the colliding polarities and explored the dependence on collision angle, speed and coherence of spots during collision, it did not account for the (at minimum) quadrupolar topology such complex ARs would exhibit on the Sun. Nonetheless, they found that the interaction of small sunspots with about $3\times 10^{21}$~Mx can lead to flares releasing about $2\times 10^{31}$~erg, corresponding to lower M-class flares. Furthermore, these flares were very efficient and released about $50\%$ of the free energy stored in the corona. Whether the collisional shearing led to a pre-eruption MFR was dependent on the coherence of the colliding sunspots, as shown in the parametric study of \citet{Rempel:etal:2023}. In fact, a pre-eruption MFR was found in a setup with less coherent sunspots where the cPIL was more persistent in time (longer lasting).

For situations with a pre-existing MFR, the torus instability \citep{Kliem:Torok:2006} is a possible path to eruption. In this situation the MFR has to rise to a height where the decay index of the external strapping field is large enough to allow for the MHD instability to set in. While the instability is an ideal MHD instability, the processes that build the MFR and allow for its rise eruption typically do involve reconnection. This process was studied by \citet{Aulanier:etal:2010} using zero beta simulations, driven at the bottom boundary by a combination of line-tied shearing motions and flux cancellation due to magnetic field diffusion. The formation of the MFR starts with bald-patch and subsequently the formation of a hyperbolic flux tube leading to a slow rise of the growing MFR until it reaches the threshold for torus instability \citep[see also][for more recent work]{Cheng:etal:2023:slow-rise,Xing:etal:2024:slow-rise}.   

In this publication we report on a generalization of the setup from \citet{Rempel:etal:2023}. We start from a quadrupolar configuration that mimics AR 11158 in terms of spot sizes and flux content and construct the sub-photospheric footpoint driving such that it imposes an amount of collisional shearing as found in observations. This study does not aim at reproducing AR 11158 in detail as we do \emph{intentionally neglect flux emergence} as it is done in \emph{data-driven} simulations such as \citet{Fan:2024:AR11158,Chen:2026:AR11158}. This study is a \emph{data-inspired} numerical experiment that aims at quantifying the role of collisional shearing (most notably, the characteristic proper motions and collision/flux cancellation of magnetic polarities) in separation from other processes, such as active flux emergence.

In Section \ref{sec:sim} we provide a detailed description of the model setup, in Section \ref{sec:results}  we present the results with special emphasis on the evolution above the cPIL that leads to the formation of a MFR and triggering of a X-flare, the conclusions are given in Section \ref{sec:concl}.

\section{Simulation Setup} \label{sec:sim}
\begin{figure*}
    \centering
    \resizebox{\hsize}{!}{\includegraphics{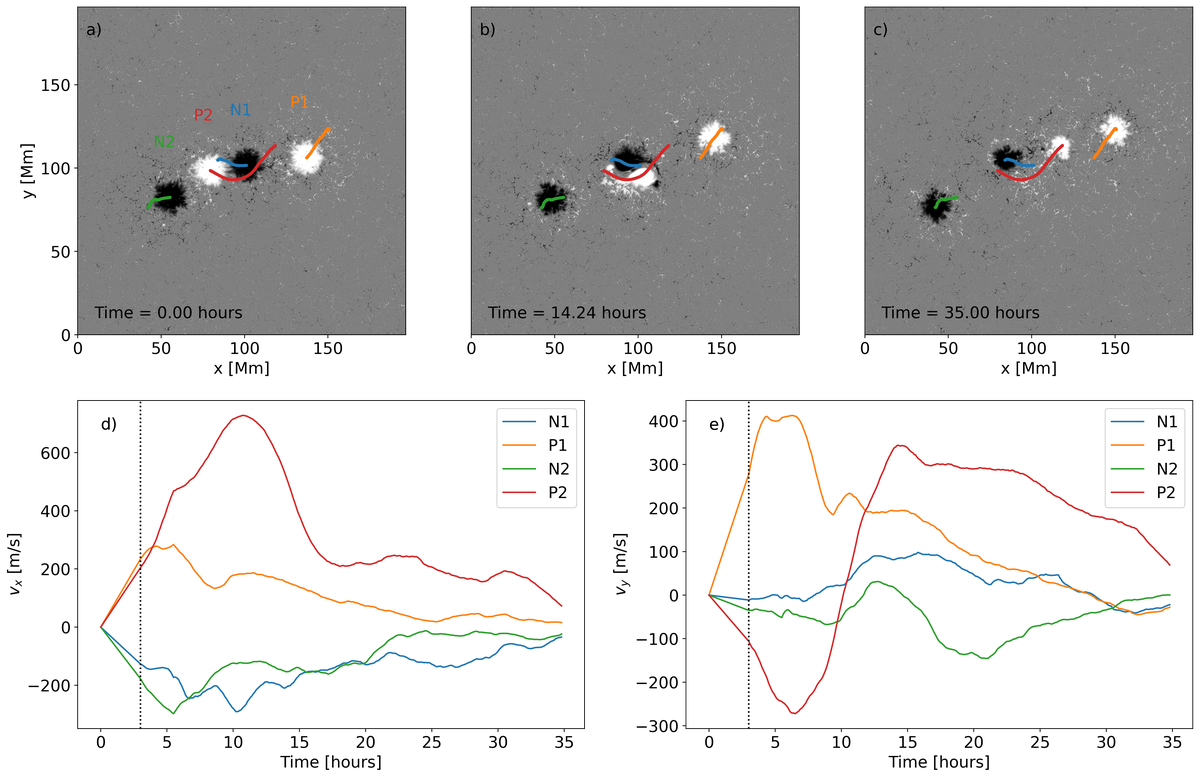}}
    \caption{Spot tracks used in the data-inspired simulation. Top panels show magnetograms with overlaid spot tracks for a) the initial state, b) during the time collisional shearing produces an X-flare and c) the end of the simulation run. Bottom panels show the corresponding velocities in the horizontal x-direction (d) and y-direction (e). Note that the simulation was sped up by a factor of 2 as described in the text, the velocities shown are increased by a factor of 2 compared to those extracted from observations.}
    \label{fig1}
\end{figure*}
We use the coronal extension of the MURaM radiative MHD code \citep{Rempel:2017:corona} that has been ported to GPUs \citep{wright:etal:2021,wright:etal:2023} for the numerical experiments reported here. This work is an extension of the bipolar colliding spot simulation from \citet{Rempel:etal:2023} in two ways: 1.) we generalize the setup to a more realistic quadrupolar configuration and 2.) we mimic the collisional shearing observed in AR 11158 by using the evolution of the observed sunspot centroid positions for the footpoint driving. To this end we consider the time frame in the evolution of AR 11158 from 2011-2-13 8:00 to 2011-2-15 24:00, comprising 62 hours of its evolution. This phase focuses on the time after the initial flux emergence, when collisional shearing due to the moving and colliding sunspots is most prominent \citep[see, e.g. Figure 3 in ][]{Chintzoglou:etal:2019}. As discussed further below, we do accelerate the time evolution by a factor of 2.

We use a numerical simulation domain with an extent of $196.608\times 196.608\times 110.592\,\mbox{Mm}^3$ with a grid spacing of $192\times 192\times 64 \, \mbox{km}^3$, corresponding to a domain with $1024 \times 1024\times 1728$ grid cells. In the vertical direction the domain extends from about 7.5 Mm beneath to 103 Mm above the photosphere. The side boundaries (x and y directions) are periodic. 

The magnetic field at the top boundary is matched to a potential field. The position of the top boundary is with about $100$~Mm high enough that this choice does not impact the pre-flare energy buildup. However, CME eruptions that reach the top boundary will be impacted as the associated currents cannot pass through the boundary. They are dissipated in a thin layer with enhanced numerical resistivity near the boundary and the associated heating is removed as it is non-physical. We conducted two simulations, one with a top boundary closed for flows ``closed simulation'', and one with a top boundary that is open for upflows during the strongest flares that also lead to CME eruptions ``open simulation''. In this paper we focus the discussion on the ``open" simulation, which is more realistic by allowing CMEs to remove mass and energy from the simulation domain.   

For the bottom boundary 7.5 Mm beneath the photosphere we use the open boundary description from \citep{Rempel:2014:SSD} outside the regions that correspond to the footpoints of the four sunspots. At the footpoints of the sunspots we impose motions following an approach that builds on \citet{Rempel:etal:2023}. While \citet{Rempel:etal:2023} only imposed the footpoint motions in the two boundary layers of the code, in this paper we also impose motions in the lowermost 3 domain cells in order to have a stronger forcing of the imposed motions. In the following we follow here the convention that the indices $-2$ and $-1$ correspond to the boundary cells, whereas the lowermost domain cells are $0$, $1$ and $2$, the physical boundary is located in-between the grid levels $-1$ and $0$.

The footpoints of the spots are defined through a combination of geometrical and field strength based criteria. For the 4 polarities in our setup we have the positions $[x_i(t),y_i(t)] _{i=1\ldots 4}$ as function of time. We consider a circle $\sqrt{(x-x_i(t))^2+(y-y_i(t))^2}<R_{\rm BND}=10$~Mm around each centroid position in which we implement our forcing terms if the magnetic field strength is large enough within this circle. To this end we define the following field strength based function ($B_{\rm BND}=2.5$~kG):
\begin{equation}
	f_B = \frac{{B_z}_0^4}{B_{\rm BND}^4+{B_z}_0^4}
\end{equation}
For the momentum $\vec{M}$ we use the following forcing ($\varrho$ denotes the mass density, the values in the boundary cells follow from Eq. (\ref{rho_bnd})):
\begin{eqnarray}
    \vec{M}_{2}&=&\vec{M}_2 \,a_{M}+\varrho_{2}\vec{v}_{\rm BND}\,(1-a_M)\nonumber\\
    \vec{M}_{1}&=&\vec{M}_1 \,a_{M}+\varrho_{1}\vec{v}_{\rm BND}\,(1-a_M)\nonumber\\
    \vec{M}_{0}&=&\vec{M}_0 \,a_{M}+\varrho_{0}\vec{v}_{\rm BND}\,(1-a_M)\nonumber\\
    \vec{M}_{-1}&=&\vec{M}_0 \,a_{M}+\varrho_{-1}\vec{v}_{\rm BND}\,(1-a_M)\nonumber\\
    \vec{M}_{-2}&=&\vec{M}_1 \,a_{M}+\varrho_{-2}\vec{v}_{\rm BND}\,(1-a_M)\label{Mbnd}\\
\end{eqnarray}
with
\begin{equation}
	a_M=1-0.2\, f_B\,,
\end{equation}
and
\begin{equation}
	\vec{v}_{\rm BND}=[\dot{x}_i(t),\dot{y}_i(t),0].
\end{equation}
For weak magnetic field ($f_B\ll 1$) we have $a_M\rightarrow 1$, which implements a symmetric mass flux boundary condition with no driving. For strong magnetic field ($f_B=1$) the boundary velocity is enforced with a coupling parameter of $0.2$, i.e. the boundary values are $80\%$ derived from a symmetic extrapolation and to $20\%$ given by the forcing term. This ''soft" coupling was found to minimize numerical artifacts, while still strongly forcing the desired flow pattern.  For the magnetic field $\vec{B}$ we use the boundary condition:    
\begin{eqnarray}    
    \vec{B}_{-1}&=&\vec{B}_0(1-f_B)  + \vec{B}_0 \,a_{B} \,f_B \nonumber\\
    \vec{B}_{-2}&=&\vec{B}_1(1-f_B)  + \vec{B}_0 \,a_{B}^2 \,f_B\label{Bbnd}\\
\end{eqnarray} 
with 
\begin{equation}
	a_B=\sqrt{\vert\vec{B}_0\vert / \vert\vec{B}_2\vert}\;_{>0.8}^{<1.2}\,.
\end{equation} 
For weak magnetic fields ($f_B\ll 1$) this implements a symmetric boundary condition for all magnetic field components as used for small-scale dynamos in \citet{Rempel:2014:SSD}, for strong field the magnetic field is extrapolated  into the boundary cells using the magnetic field gradient near the bottom boundary. We limit the parameter $a_B$ between values of $0.8$ and $1.2$ to prevent numerical instabilities. Finally, the gas pressure $p$ in the boundary cells is computed as: 
\begin{eqnarray}            
    p_{-1}&=&\bar{p}_0\,a_{p}+p_0^{\prime}\,(f_B+(1-f_B)\,c_{\rm dmp})\nonumber\\
    p_{-2}&=&\bar{p}_0\,a_{p}^2+p_1^{\prime}\,(f_B+(1-f_B)\,c_{\rm dmp})\label{pbnd}
\end{eqnarray}
with
\begin{equation}
	a_p=\frac{p_{\rm BND}}{\sqrt{\bar{p}_0\,\bar{p}_1}}
\end{equation}
Here $\bar{p}$ denotes the horizontally averaged pressure and $p^{\prime}$ the pressure perturbation. $p_{\rm BND}$ is the value of the boundary pressure. In strong field regions we simply add the symmetric pressure perturbations, while they are damped ($c_{\rm dmp}=0.8$) in weak field regions similar to the treatment in \citet{Rempel:2014:SSD} (without damping we found that strong standing oscillations can develop). For the entropy in the boundary cells we use a symmetric condition in outflow regions and impose a fixed value everywhere else (upflows, footpoints of sunspots), chosen such that radiative losses in the quiet Sun photosphere lead to a solar energy flux of about $6.3\times 10^{10}\mbox{erg}\,\mbox{cm}^{-2}\,\mbox{s}^{-1}$. The density and internal energy in the boundary cells follows from the equation of state:
\begin{eqnarray}
	\varrho_{-1}&=&\varrho(p_{-1},s_{-1})\nonumber\\	\varrho_{-2}&=&\varrho(p_{-2},s_{-2})\label{rho_bnd}\\
	(E_{\rm int})_{-1}&=&E_{\rm int}(p_{-1},s_{-1})\nonumber\\	
    (E_{\rm int})_{-2}&=&E_{\rm int}(p_{-2},s_{-2})\label{eint_bnd}  
\end{eqnarray}
The equation of state tables are based on the FreeEos package \citep{Irwin:2012:freeeos} and use Asplund 2009 abundances \citep{Asplund:etal:2009}. For the optically thin radiative loss in the corona we use a loss function computed with Chianti 10.1 \citep{Dere:2023:Chianti} using coronal abundances from \citet{Feldman:1992}. For computation of synthetic observables we use AIA and GOES 18 response functions based on the same choice of abundances.

We use the positions $[x_i(t=0),y_i(t=0)] _{i=1\ldots 4}$ to initialize our domain with a quadrupolar setup. We choose an initial magnetic field structure following the sunspot simulations of \citet{Rempel:2012:penumbra}, as detailed in the Appendix A therein. We used a Gaussian profile for the radial and an exponential profile for the vertical field dependence. Magnetic field strength $B_0$ and radius $R_0$ at the bottom boundary are $10$~kG and $5.05$~Mm, which leads to a magnetic flux of initially $8\cdot 10^{21}$~Mx for each spot. We initially relaxed the sunspots in a photospheric MHD simulation of $8.192$ Mm depth, the parameter $z_0$ was chosen such that the initial field strength was $2$~kG at the top boundary of the photospheric simulation. We ran this setup for 4 hours of solar time until the spots were thermally and dynamically relaxed. After this initial relaxation the flux remaining in concentrated spots was about $6\cdot 10^{21}$~Mx comparable AR 11158. As a next step we added the corona by initializing the magnetic field above the photosphere with a potential field and initializing the density and internal energy with a mean stratification taken from the quiet Sun simulation of \citet{Chen:etal:2022}. This setup was evolved for about 3 hours of solar time to allow the coronal part of the simulation to reach a relaxed state. Figure \ref{fig1}a) shows the vertical magnetic field in the photosphere (on a geometric height corresponding to the average optical depth $\tau_{500}=1$) for this initial state. The overlaid colored lines do show the sunspot centroid trajectories we use for the boundary forcing. Panel b) shows the configuration after $14.24$ hours when the central polarities are colliding and, as we discuss later, produce an X-flare. Panel c) shows the configuration at the end of the simulation. Panel d) and e) show the horizontal velocities $\dot{x}_i(t)$ and $\dot{y}_i(t)$. We initially added a linear ramp-up phase of 3 hours during which the velocities increase linearly from zero to the centroid speed of our first data point (corresponding to 2011-2-13 8:00). The time evolution is accelerated by a factor of 2, leading to velocities as high as 730 m/s and shortening the actual data-driven duration of the simulation to 32 hours. This was done for 2 reasons: 1.) to save computing time and 2.) to counter sunspot decay that is faster in the simulation than in the observed active region. We performed test runs (only using the photospheric part of the domain) and found that with the driving at the original speed, the spots would have mostly decayed after 64 hours of evolution.

The numerical setup described here is a ``data-inspired" simulation, i.e. we use a setup that possesses a similar amount of collisional shearing as found in AR 11158. However, since we do neglect sunspot evolution driven by ongoing flux emergence there is no expectation that this simulation can capture specific details of AR 11158. The main purpose is to study how efficient collisional shearing is in energizing the corona and in driving flares and CMEs when scaled up to the size of a solar active region.

\section{Results} 
\label{sec:results}
\subsection{General description of evolution}
\label{sec:general-prop}

\begin{figure*}
    \centering
    \resizebox{\hsize}{!}{\includegraphics{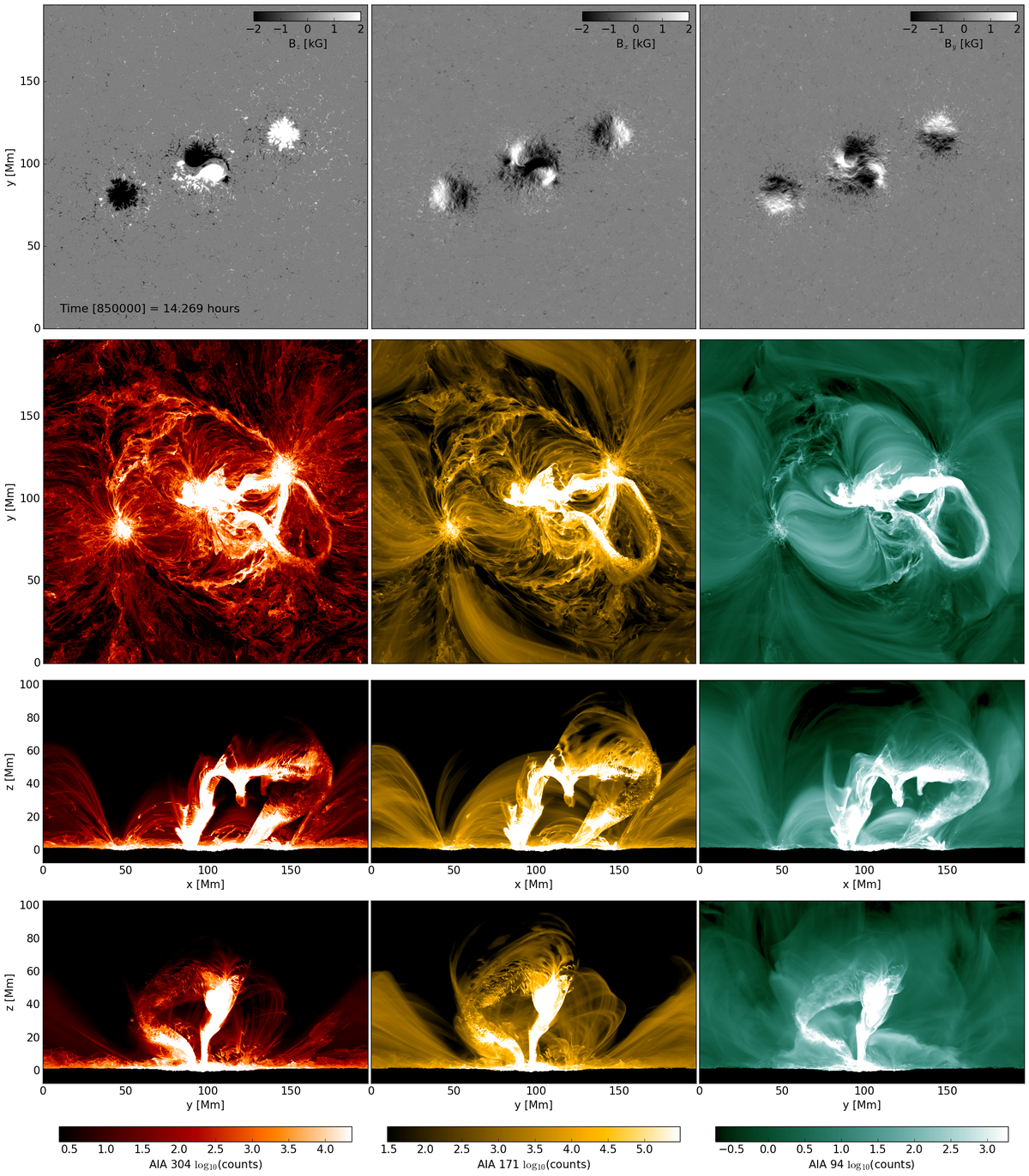}}
    \caption{Top row: Photospheric ($\tau=0.1$) magnetic field: $B_z$, $B_x$, $B_y$. Second row: Synthetic AIA emission in the $304$, $171$, and $94$ passbands for a top view. Third and fourth rows: AIA emission for view along y-axis and x-axis, respectively. We show a snapshot at $t=14.269$ hours, which shows the ejection of a CME  following an X-flare. See the animation for the full time evolution of the simulation.} 
    \label{fig2}
\end{figure*}

Figure \ref{fig2} and the associated animation show the evolution of the simulation in terms of photospheric magnetic field (extracted on the $\tau_{500}=0.1$ surface) and synthetic coronal emission computed for the $304$\AA, $171$\AA, and $94$\AA\, AIA channels from 3 view points along the Cartesian coordinate directions. The movie starts with the initial potential magnetic field configuration after it had been dynamically relaxed for frozen in positions of the sunspots. This relaxation led to a stable self-maintained corona (i.e. a balance between the Poynting flux resulting from photospheric magnetoconvection and radiative losses in the corona). In the early stages of the simulation there are clear signs of coronal rain forming. With the start of the collision between the polarities N1 and P2 we see an increase in the AIA $94$\AA\, counts which initially peaks around $t=10$~hours. Strongly sheared magnetic fields are manifest in the following 4 hours through sigmoidal AIA emission structures. As we show later, a low-lying magnetic flux rope (MFR) is forming during this time. At $t=14.26$ hours this flux rope is erupting in an X-flare with associated coronal mass ejection, releasing about $2.2\cdot 10^{32}$ ergs of energy. The coronal volume is heated during this process to temperatures exceeding the range of the displayed AIA channels. As a consequence, the EUV emission is pushed mostly towards the lowermost 1/3 of the simulation volume. As the corona cools, loop systems visible in AIA $171$\AA\, do expand upwards, while strong coronal rain is visible in AIA $304$\AA. The emission in AIA $94$\AA\, remains more diffuse for the remainder of the simulation. As the polarities N1 and P2 continue passing each other we find a series of $10$ more flares in the B-M range with an additional $3$ CME eruptions.  

\begin{figure*}
    \centering
    \resizebox{\hsize}{!}{\includegraphics{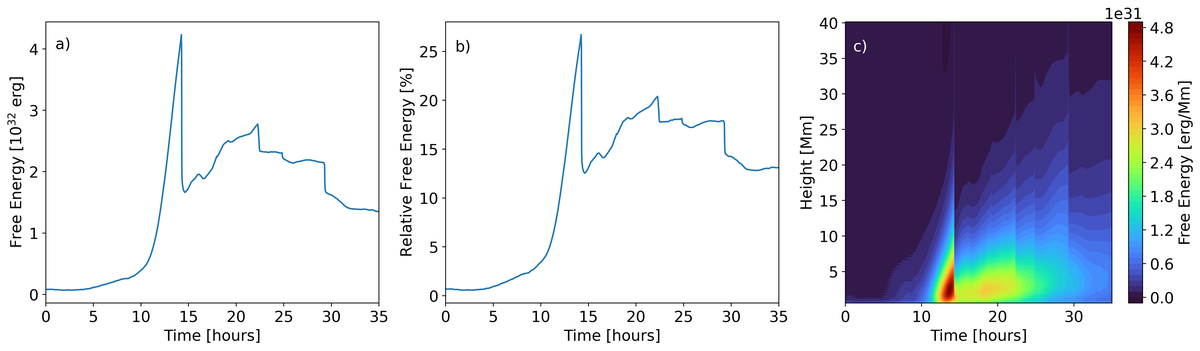}}
    \caption{Temporal evolution of free magnetic energy in the simulations. a) Evolution of the free energy. b) Evolution of free energy relative to the total magnetic energy. c) Evolution of free energy as function of height and time. Note that most of the free energy, in particular the free energy leading to the X-flare is stored below a height of 5-10~Mm.} 
    \label{fig3}
\end{figure*}

Figure \ref{fig3} shows the evolution of free magnetic energy in the corona. Since the simulation is computed with horizontally periodic boundary conditions, the free magnetic energy is defined with respect to the periodic potential field solution, that only depends on the photospheric magnetic field (computed through the Fourier transform method). Panel a) show the free energy defined as:
\begin{equation}
    E_f(t)=\int \frac{B_{\rm MHD}^2(x,y,z,t)-B_{\rm pot}^2(x,y,z,t)}{8\pi} dV
\end{equation}
where the volume V is the full extent of the coronal volume of the simulation domain starting from a constant height surface corresponding to the average height of $\tau_{500}=1$. Panel b) shows the relative free energy. The buildup of energy starts around $10$ hours, when the polarities N1 and P2 start to shear off against each other. The free energy increases to more than $4\cdot 10^{32}$ erg ($25\%$ relative free energy), out of which around $50\%$ are released in the X-flare. A similar value in ``flare efficiency" was also found in \citet{Rempel:etal:2023} in a smaller-scale bipolar collisional shearing setup. The flare efficiency for the subsequent flares is in the $5-20\%$ range. Panels c) and d) show the height distribution of free magnetic energy (i.e. the horizontal integral of $E_{\rm mag} -E_{\rm pot}$) as function of time:
\begin{equation}
\langle E_f(z,t)\rangle_{h}=\int_0^{L_x}\int_0^{L_y} dx dy \frac{B_{\rm MHD}^2(x,y,z,t)-B_{\rm pot}^2(x,y,z,t)}{8\pi}
\end{equation}
Here $L_x$ and $L_y$ are the horizontal extents of the simulation domain. Most of the free magnetic energy is stored in the lowermost $5-10$~Mm above the photosphere. 

\begin{figure*}
    \centering
    \resizebox{\hsize}{!}{\includegraphics{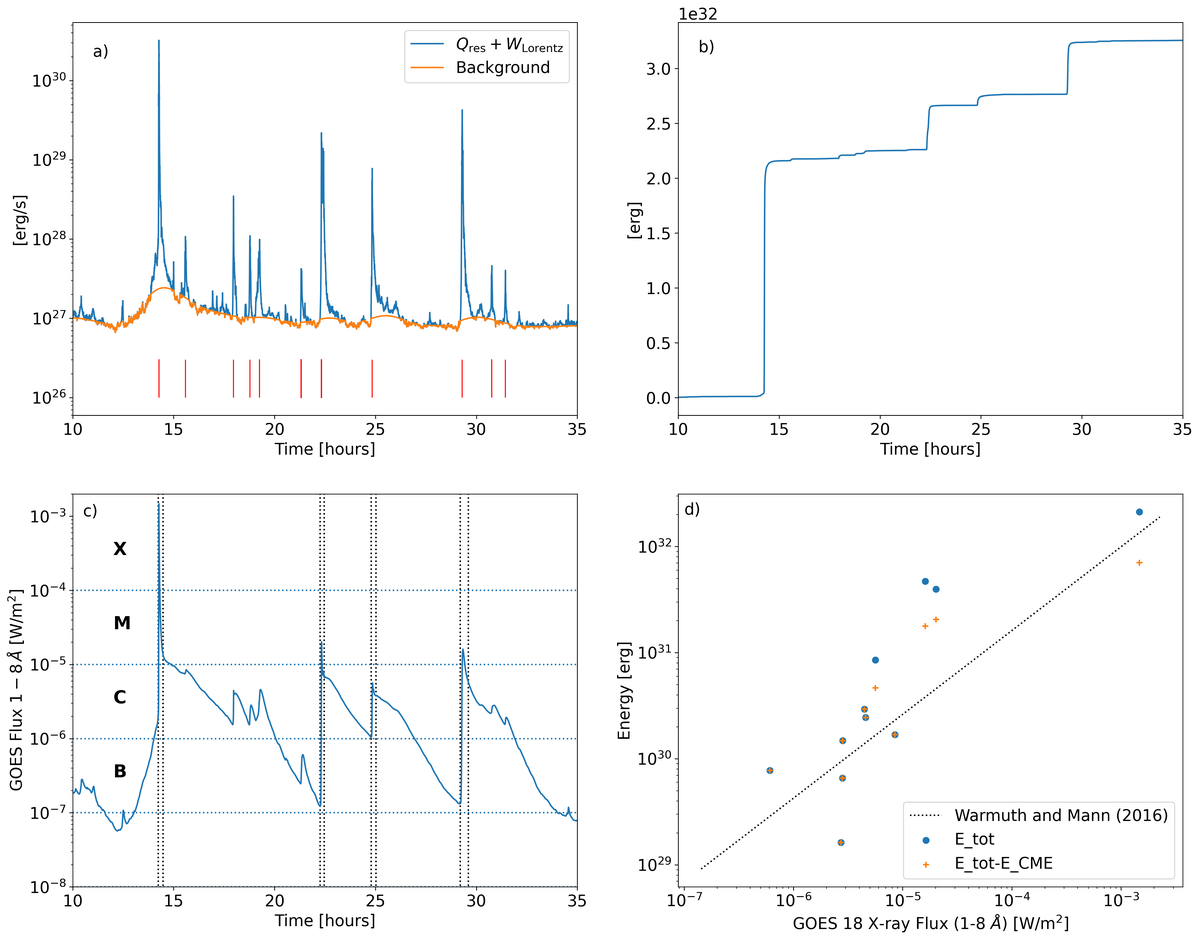}}
    \caption{Energy released and relation to GOES flux. Panel a): Sum of Lorentz force work and resistive heating, the orange line indicates a background we subtracted for determining the flare energy. Vertical red lines indicate flares studied further. Panel b): Total cumulative energy released in flares. Panel c): Synthetic GOES 18 X-ray flux. With dotted vertical lines we indicate times during which the top boundary was open for CMEs to leave. Panel d) Relation between GOES 18 flux and flare energy. Blue dots show the total released energy, orange crosses the thermal energy of the flares after subtracting the energy the resulting CMEs transport through the open top boundary of the simulation. The dotted line indicates the relation that was inferred from observation by \citet{Warmuth:Mann:2016} (since this analysis was based on earlier GOES data, we shifted the curve by a factor of $1.42$ for consistency with GOES 18).}
    \label{fig4}
\end{figure*}

In Figure \ref{fig4} we show the cumulative energy released in the sequence of flares. To this end we consider the sum of Lorentz-force work and numerical resistive heating and subtract the non-flaring background that corresponds to coronal heating (panel a). While the X-flare releases about $2.16\cdot 10^{32}$ erg of energy, the sequence of weaker flares adds another $1.1\cdot 10^{32}$ erg (panel b). Panel c) shows the synthetic GOES 18 $1-8$\AA\, X-ray flux computed from the simulation. The $4$ strongest flares do have associated CMEs and the vertical dotted lines indicate the times during which the top boundary was open to allow for the CMEs to leave the domain. Panel d) shows the relation between synthetic GOES flux and the released energy for the flares in indicated in panel a) by red vertical lines. These flares have in common that they do have distinct GOES peaks. Blue dots show the total energy released, orange crosses the thermal energy remaining in the volume after subtracting the thermal, kinetic and magnetic energy of the CMEs leaving the domain (quantified by integrating the energy flux at the top boundary of the domain over the horizontal extent and in time). The dotted line indicates the observations based relation from \citet{Warmuth:Mann:2016}. While the simulated flares do show generally a similar trend, we note that the GOES class of the simulated flares does depend on the amount of chromospheric evaporation during flares. That does depend on the treatment of the transition region \citep[e.g.][]{Johnston:2021:TRAC} as well as the amount of non-thermal energy that is transported into the chromosphere and emitted there through non-EUV processes. These details need to be taken care off before a detailed comparison with an observed relation is well founded -- in this paper we use the GOES class primarily to classify the simulated flares in relation to each other. 

Both aspects are being currently improved upon and will be presented in future publications. Test simulations with both an implementation of the TRAC method \citep{Johnston:2021:TRAC} and a separation of non-thermal energy, including deposition below the transition region do generally lead to similar GOES classes as both effects partially compensate each other (the TRAC method enhances chromospheric evaporation, while taking out non-thermal energy and depositing that energy below the transition region does reduce chromospheric evaporation).

\subsection{Flare durations}
\label{sec:duration}

\begin{figure*}
    \centering
    \resizebox{0.9\hsize}{!}{\includegraphics{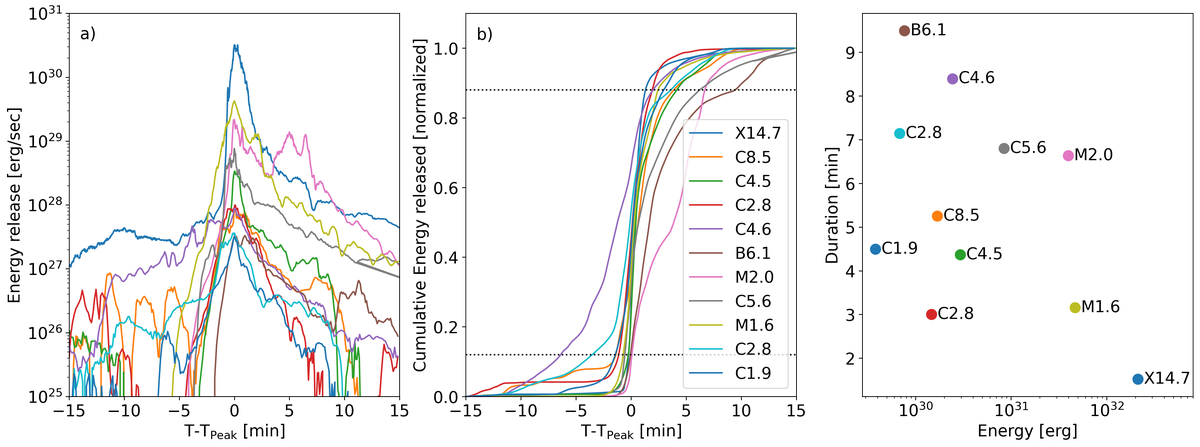}}
    \caption{Duration of the flares based on the volume integrated coronal energy release (sum of Lorentz force work and resistive heating). Panel a) All flares are stacked together based on the energy release peak as zero point in the time axis. Panel b) normalized integrated energy release rates. Panel c) Flare duration based on $12\%/88\%$ thresholds in panel b). }
    \label{fig5}
\end{figure*}

Another well studied aspect of flares is their duration. \citet{Reep:2019:FlareDuration} studied the duration of flares in the C-X range by analyzing the FWHM of GOES X-ray emission. They found that the mean and median flare durations do not vary strongly with flare class and are around 10-13 minutes, while for each flare class the duration has a wide spread ranging from about a minute to an hour. While \citet{Reep:2019:FlareDuration} based their analysis on the width of GOES emission peaks, we use in our simulations directly the total energy release in the corona (sum of Lorentz force work and numerical resistive heating). The reason is that many of the weaker flares are found in the declining phase of stronger earlier flares, which leads to less pronounced GOES peaks, while the coronal energy release does have a larger prominence. In Figure \ref{fig5}a) we show the energy release curves for all flares stacked with respect to the peak time, panel b) shows the cumulative energy release normalized to one for all flares. If the the shape of the energy  release would be Gaussian, the curves in panel b) would correspond to $\frac{1}{2}\,(1+\erf(t))$, with $\erf$ denoting the error function. The FWHM corresponds the to the 0.12 and 0.88 thresholds of the cumulative energy release, which we use to define the flare duration as shown in panel c). Overall we find flare durations in $1-10$ minutes range, which populate only the lower half of the distributions found in \citet{Reep:2019:FlareDuration}. In particular the X14.7 flare is with $1.5$ Minutes exceptionally short. While it is expected that our measure based on the coronal energy release may be shorter compared to the GOES curve FWHM and perhaps more correlated to the rise of the GOES emission \citep{Neupert:1968}, this also suggests that the global flare reconnection rate found in our simulations is up to one order of magnitude faster than the average solar flare. While the local reconnection rate in the flare current sheets does reach values of about $R=0.1$ (ratio of reconnection inflow speed to Alfv{\'e}n velocity) as expected for fast reconnection (the reconnection is due to numerical diffusivities, which lead due to their non-linearity always to fast reconnection), the global reconnection rate appears to be identical to the local rate.
One possible reason for that is that due to limited numerical resolution the aspect ratio of the current sheet is too low to allow for fragmentation, i.e. multiple X-line reconnection as it would be expected for the solar parameter regime \citep{Ji:etal:2022}. Under such conditions the global reconnection rate can drop to $R=0.01$ \citep{Uzdensky:2010:ReconnectionPlasmoid,Huang:Bhattacharjee:2010:HighLundquist,Ji:etal:2022}, which would lead to flare durations comparable to the upper half of the observed solar distribution. For the following discussion it is critical to keep in mind the short duration of the simulated flares, as this leads to rather fast propagation of flare ribbons, ribbon derived reconnection rates, high energy fluxes into the flare ribbons and generally fast flare induced flows.

\subsection{Detailed analysis of the X-Flare}
We focus the following discussion on the evolution that leads to the X-flare, which is the first flare in the series of more than 10 flares present in this simulation. We do not analyze the conditions leading to the other flares in this publication.

\subsubsection{Formation of pre-eruption flux rope}
\begin{figure*}
    \centering
    \resizebox{\hsize}{!}{\includegraphics{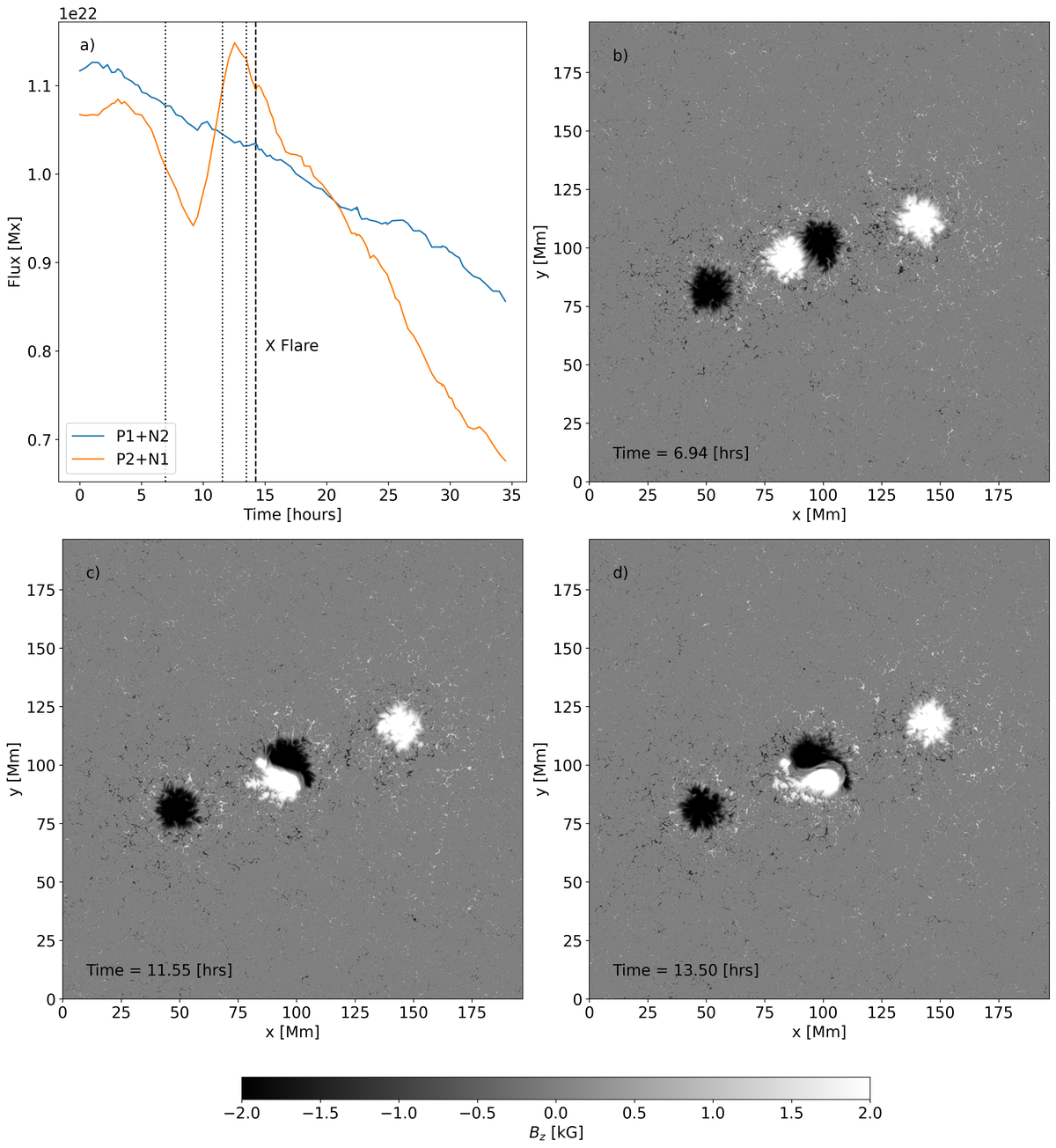}}
    \caption{Magnetic evolution of photospheric magnetic flux. Panel a) shows the unsigned flux in the outer polarities (P1+N2) in blue and the inner colliding polarities (P2+N1) in orange. The steady decline of the flux in the outer polarities indicates the level of sunspot decay in the simulation in the absence of close interaction. The evolution of the inner polarities in non-monotonic: after an initial decline during first contact the flux increases due to re-emergence of canceled flux, before it transitions to a second declining phase that contains all the flares. Dotted lines indicate representative times for which we show magnetograms in panels b), c) and d). The dashed vertical line in panel a) indicates the timing of the X-flare.}
    \label{fig6}
\end{figure*}

Figure \ref{fig6}a) shows the time evolution of the photospheric magnetic flux. We grouped the spots P1/N2 (outer, non-interacting polarities) and P2/N1 (inner, interacting polarities) together and display the unsigned magnetic flux. The outer polarities show a steady decay of about $1.85\cdot 10^{21}$~Mx/day for both spots combined. This rate is comparable to what was found by \citet{Rempel:2015:moat} for the decay of a ``naked" spot in simulations (about $1\cdot 10^{21}$~Mx/day for a single spot with $5\cdot 10^{21}$~Mx flux). \citet{Rempel:2015:moat} also found that spots with a penumbra are more stable -- a penumbra cannot be maintained in the simulations at hand due to lack of resolution. We note that this unavoidable decay was our main reason for speeding up the time evolution by a factor of 2 compared to AR 11158. We also note that the initial combined flux of the inner polarities is lower than the flux of the outer polarities. The time $t=0$~hours corresponds to the time when we start the footpoint driving as described in Figure \ref{fig1}. This time was preceded by $4$~hours of photospheric relaxation and $3$~hours of coronal relaxation during which the two inner polarities decayed more rapidly than the outer polarities due to their close proximity. Initially at $t=-7$~hours all polarities had the same amount of flux. 

The evolution of unsigned flux for the inner polarities in non-monotonic. The magnetic flux drops quickly during the initial approach of the polarities, a snapshot of that phase is shown in panel b) (corresponding to the first dotted line in panel a). Starting from about $t=9.5$~hours this drop changes to a rapid increase during the time with the strongest shear when P2 and N1 pass each other. This is due to re-emergence of previously canceled and submerged flux. This re-emergence is better visible in this simulation due to the absence of flux emergence, that would at least partially mask this process in active regions such as AR 11158. A representative snapshot is displayed in panel c), corresponding to the second dotted line in panel a). This phase lasts to about $t=12$~hours (about 2 hours before the X-flare), when the combined flux of the inner polarities reaches a value that even exceeds the flux at $t=0$~hours (although still lower than the starting value at the initialization at $t=-7$~hours). After $t=12$~hours the flux decays monotonically until the end of the simulation. Panel d) shows a magnetogram about $45$~minutes before the X-flare. Unlike the magnetogram at $t=11.55$~hours  (panel c), the polarities are more separated at the cPIL due to more complex processes of subduction and re-emergence of flux. We also highlight the feature at the leading edge of the fastest moving positive polarity (P2), where there is a small band of negative flux. This negative flux is disappearing as P2 continues to move to the right and it is this area above which a pre-eruption MFR is forming and the X-flare is triggered. 

\begin{figure*}
    \centering
    \resizebox{\hsize}{!}{\includegraphics{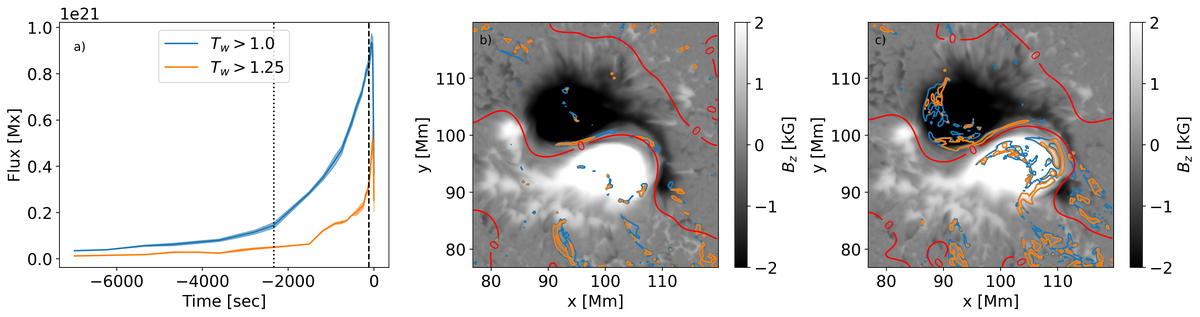}}
    \caption{a) Time evolution of magnetic flux contained in the footpoints of field lines with twist numbers larger than 1 (blue) and 1.25 (orange). The shaded area around each line indicates the range of uncertainty estimated based on values computed from the positive and negative footpoint, respectively. The vertical dotted (dashed) line indicates the time for which we show magnetograms in panels b) and c), respectively. Blue and orange contours in panels b) and c) indicate the footpoints of field lines with twist numbers larger than 1 (blue) and 1.25 (orange). The red contours indicate the polarity inversion lines.}
    \label{fig7}
\end{figure*}

Figure \ref{fig7} shows the evolution of magnetic flux in regions with strongly twisted magnetic field. We used the GPU version of the FastQSL code \citep{Zang:2022:FastQSL} to compute the twist number ($T_w$) of field lines in the coronal volume of the simulation, which is defined as:
\begin{equation}
    T_w=\frac{1}{4\,\pi}\int\frac{(\vec{\nabla}\times\vec{B})\cdot\vec{B}}{\vert\vec{B}\vert^2}
\end{equation}
The integration was started from a constant height surface located about $192$~km (3 grid points) above the average height of the $\tau_{\rm 500}=1$ surface. We compute the flux above $T_w$ thresholds of $1$ and $1.25$ for both polarities and display the average in panel a). The shaded area about these curves indicates the spread in the results from the positive and negative footpoints alone, indicating overall very consistent results. A notable buildup of a MFR starts around $1$~hour before the X-flare and in particular the flux in regions with $T_w>1.25$ increases rapidly just minutes before the flare. The panels b) and c) show the photospheric magnetogram with overlaid contours of $T_w=1$ and $T_w=1.25$ as well as the PIL (in red), the latter being based on a magnetogram smoothed with Gaussian (FWHM=$4.5$ Mm). Overall the pre-eruption MFR reaches a flux of about $10^{21}$ Mx ($T_w>1$) or $5\cdot 10^{20}$ Mx ($T_w>1.25$). This value is comparable to the amount of flux cancellation found prior to the eruption (see Figure \ref{fig6}). In the 2 hours before the X-flare we do find a drop of the unsigned P2+N1 flux by about $5\cdot 10^{20}$ Mx. Note that this is the drop in the net flux, after adding up both re-emergence and cancellation, so the latter is actually larger. 

\begin{figure*}
    \centering
    \resizebox{\hsize}{!}{\includegraphics{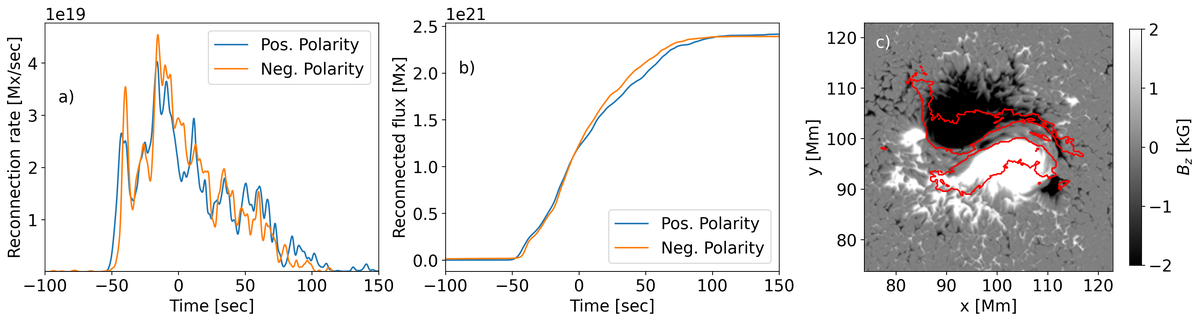}}
    \caption{a) Reconnection rate derived from progression of flare ribbons in the central region. Blue/orange lines indicate the reconnection rate derived from the positive/negative polarity flare ribbon. b) Time evolution of reconnected flux based on positive/negative polarity. c) Total area swiped by flare ribbons.}
    \label{fig8}
\end{figure*}

Figure \ref{fig8} quantifies the reconnection flux based on the computation of cumulative flare ribbon maps following the approach of \citet{Kazachenko:2017:flareDB}. While flare ribbon maps in observations are typically derived from AIA $1600$\AA\, emission maps, we used the gas pressure at a height of $1$ Mm above the photosphere (average $\tau_{500}=1$) as a proxy. The deposition of heat in the flare ribbons does lead to a significant increase in pressure. Whereas typical values are around $10-20$ dyne for this height, we used a threshold of $10^3$ to identify the ribbons. Panel a) shows the reconnection rates for both the positive and negative flare ribbon fluxes, panel b) the cumulative reconnection fluxes. Panel c) shows a photospheric magnetogram with contours of the total area swept by the flare ribbons. The reconnection rate was filtered with a Gaussian (FWHM of 2 seconds) to suppress shot-term variability that would make the figure difficult to read. The reconnection flux computed from either polarity provides similar results and showing a total of reconnected flux of $2.4\cdot 10^{21}$~Mx (i.e. $4.8\cdot 10^{21}$~Mx for the total unsigned reconnection flux often used in the literature). The reconnection flux is 2.4 times larger than the flux that was present in the pre-existing MFR with $T_w>1$. Comparing this value to the flare data base of \citet{Kazachenko:2017:flareDB}, the simulated X-flare is falling into the lower range of observed X-flares, which have unsigned reconnection fluxes from $3.8\cdot 10^{21}$~Mx to $3\cdot 10^{22}$~Mx \citep[see Figure 8][]{Kazachenko:2017:flareDB}. Relative to the unsigned AR flux (see Figure \ref{fig6}), the reconnection flux is $23\%$, which falls into the middle of the distribution for X-flares \citep[see Figure 11 in][]{Kazachenko:2017:flareDB}.

The reconnection rate reaches $4.5\cdot 10^{19}$~Mx s$^{-1}$, the unsigned reconnection rate, adding both flare ribbons often used in the literature, reaches $9\cdot 10^{19}$~Mx s$^{-1}$. This value is larger by a factor of $3-10$ compared to the observed range of reconnection rates \citep[see Figure 3 in][]{Tamburri:2024}. The high reconnection rate is a direct consequence of the short flare duration as discussed in Section \ref{sec:duration}. The shorter-term variability seen in the reconnection rate is possibly related to quasi-periodic pulsations (QPPs), which will be further discussed in Section \ref{sec:QPP}.

\subsubsection{Triggering mechanism and onset of eruption}

\begin{figure*}
    \centering
    \resizebox{\hsize}{!}{\includegraphics{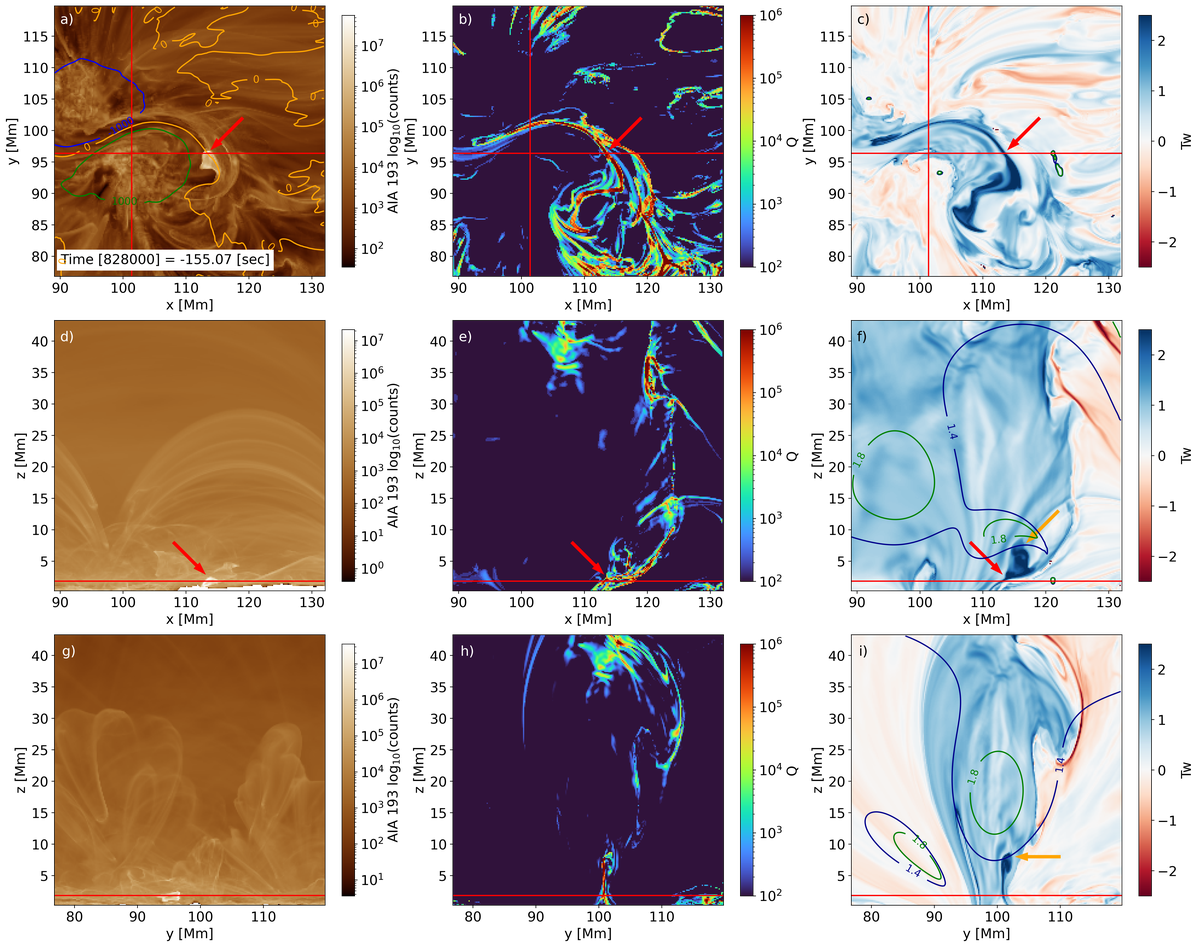}}
    \caption{Conditions during the onset of the X-flare. The left panels show synthetic AIA $193$\AA\, emission, the middle panels the field line squashing factor and the right panels the twist number. Top to bottom we show the view along the vertical, horizontal y and horizontal x directions (for AIA) and cuts for $Q$ and $T_w$ along the positions indicated by red lines. Contours in panel a) correspond to the field strengths of $B_z = [-1000,0,1000]~G$, contours in panels c),  f) and i) correspond to the decay index of $1.4$ and $1.8$. The red arrows in panels a) to f) point to the position of the first flare related brightening visible in synthetic AIA. This location corresponds to the onset of tether cutting  reconnection beneath the pre existing MFR in a height of just $\sim 2$~Mm above the photosphere. This location has a high squashing factor exceeding $10^6$. During this time the top of the MFR also passes the threshold of a decay index of $1.4$, which is indicated by the orange arrows in panels f) and i). An animation of this figure is provided.}
    \label{fig9}
\end{figure*}

\begin{figure*}
    \centering
    \resizebox{\hsize}{!}{\includegraphics{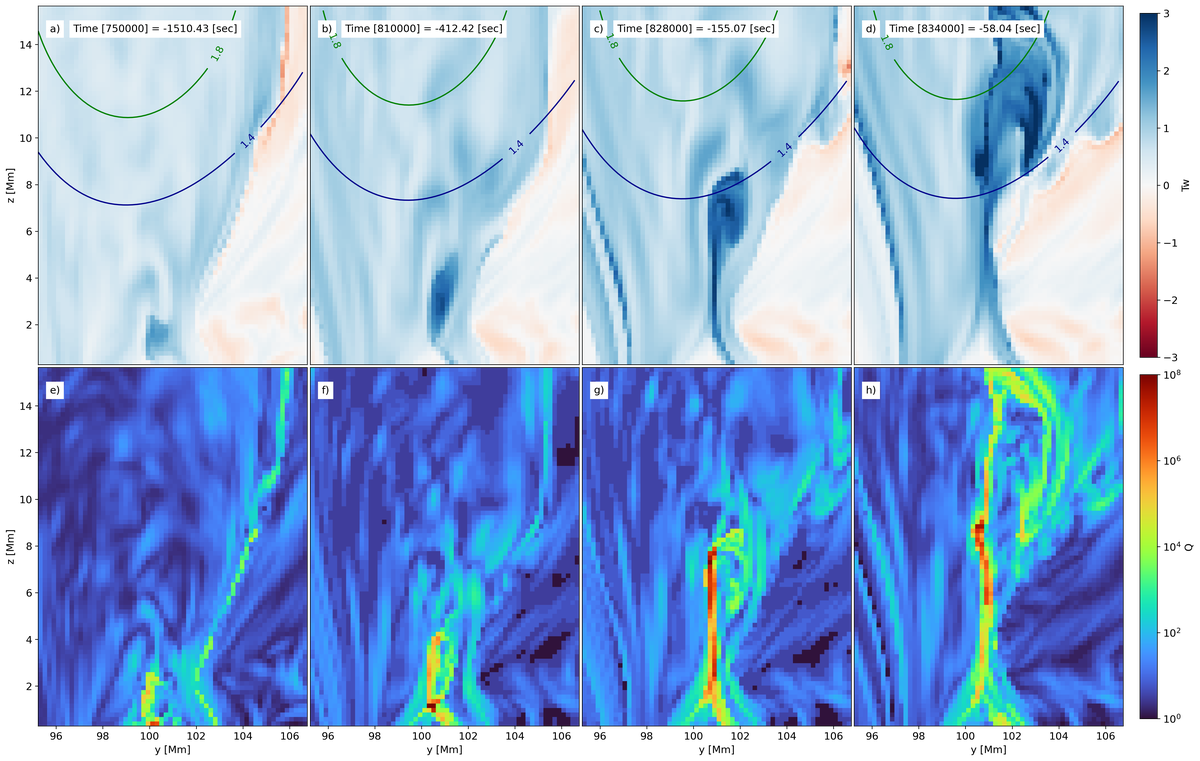}}
    \caption{Time evolution of MFR in terms of twist number (a-d) and squashing factor number (e-h) for the cross section along the y-axis in Figure \ref{fig9}.}
    \label{fig10}
\end{figure*}

Figure \ref{fig9} shows the conditions during the onset of the flare, which we define here as the time at which the first flare related EUV brightening occurs. The left panels display synthetic AIA $193$\AA\, emission seen from 3 view points. Panel a) shows the top view with contours indicating the $B_z=[-1000,0,1000]$~G levels of the photospheric magnetic field. The central panels show the field line squashing factor, the right panels the twist number. The top panels show a cut at a height of 2~Mm above the photosphere, the second row panels along the x and the bottom panels along the y-direction as indicated by the red lines in panels a-c). The red arrows in panels a) to f) point toward the first EUV brightening that is indicative of the onset of the eruption. The origin of this brightening is with a height of 2~Mm relatively low in the atmosphere and is located underneath the pre-existing flux rope as indicated in panels e) and f). At the same time, the top of the MFR is entering regions with a decay index higher than 1.4 as indicated by the orange arrow in panels f) and i). Overall, this suggests that the MFR becomes torus unstable and that tether-cutting reconnection sets in underneath the MFR which leads to the brightening in EUV emission.

The onset of torus instability and rapid reconnection is simultaneous. Figure \ref{fig10} shows the development of the MFR prior to the onset for the cross section along the y-direction in Figure \ref{fig9}. While the MFR is moving upwards towards the threshold of torus instability an elongated sheet with high squashing factor values exceeding $10^7$ is forming and reaching a length of about 8 Mm at the time of onset (panels c and g). This quasi-separatrix layer does coincide with a current sheet, which is also seen as a thin sheet in the twist number. \citet{Kliem:2014:CvsTI} demonstrated that the onset of instability and the onset of rapid reconnection due to a catastrophe are equivalent descriptions of the problem. Therefore it is not expected to see a large timing difference in the onset of instability and reconnection beneath the MFR. Figure \ref{fig10} also shows that the MFR starts forming at photospheric height, a hyperbolic flux tube (HFT) above the photosphere is present about 400 seconds before flare peak.

\begin{figure*}
    \centering
    \resizebox{\hsize}{!}{\includegraphics{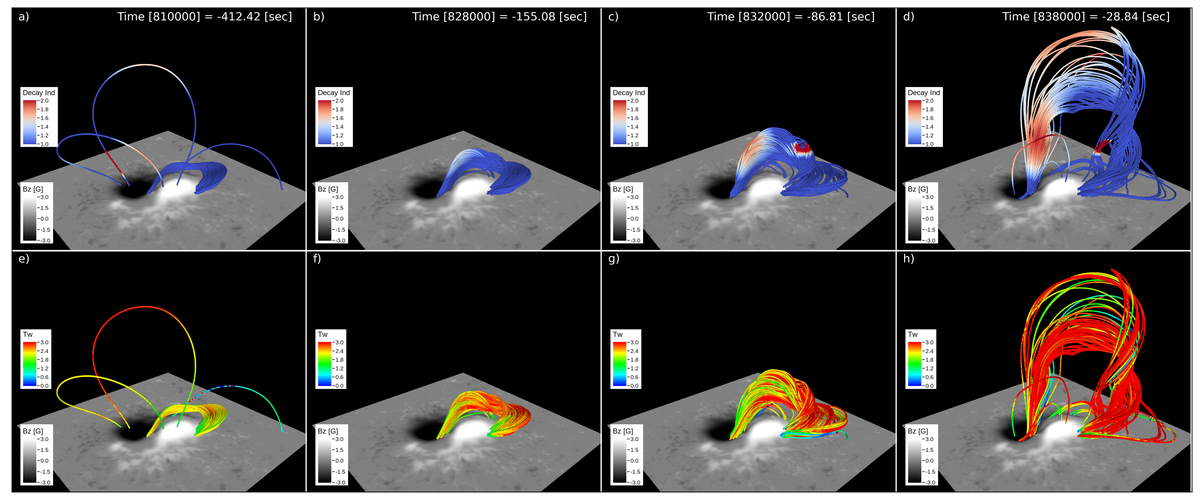}}
    \caption{Evolution of the flux rope during the onset of the eruption. Field lines were seeded based on high values of the twist number in the lowermost $6$~Mm above the photosphere. Panels a) - d) show field lines color coded by the value of the decay index, panels e) - h) by the value of the twist number. Panels a) and e) show the pre-eruption MFR with twist numbers $T_w$ around 2 in a stable configuration. Panels b) and f) correspond to the time of eruption onset also highlighted in Figure \ref{fig8}. $T_w$ reaches now values of $3$ and parts of the MFR move to heights with a decay index of $1.5$. The remaining panels show the progression of the eruption with parts of the MFR located in unstable locations. This figure was created with VAPOR \citep{2019Atmos..10..488L:vapor,sgpearse_2023_7779648}.}    
    \label{fig11}
\end{figure*}

Figure \ref{fig11} displays magnetic field lines selected based on $T_w$ covering a time range of about 6 minutes. Field lines in the top panels are color mapped to the decay index, field lines in the bottom panels are mapped to $T_w$. Note that some field lines show $T_w$ varying along their length, which is a numerical artifact caused by numerical differences in the the code we used to compute $T_w$ (FastQSL) and the code we used to display the results (Vapor). Panels a) and e) show the stable pre-eruption MFR. The MFR is located in regions with a decay index below $1.5$ and has moderate $T_w$ of mostly less than $2$. At the time when the first EUV signature of the eruption becomes visible, panels b) and f), $T_w$ has increased to $3$ in significant parts of the MFR and the top of the MFR is starting to move into regions with a decay index around 1.5.  

\begin{figure*}
    \centering
    \resizebox{\hsize}{!}{\includegraphics{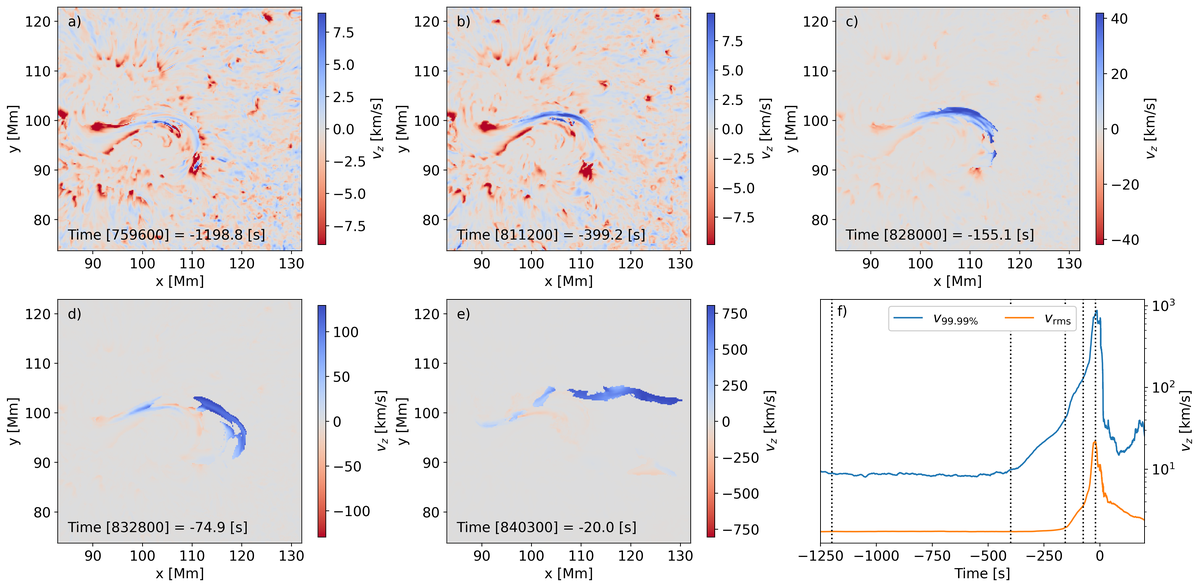}}
    \caption{Time evolution of the vertical velocity at optical depth of $10^{-4}$. We can see an accelerating rise of the pre-flare MFR. First indications of the lifting MFR are visible more than $1000$ seconds before the flare peak (panel a) and a clearly discernible upflow is present starting from $400$ seconds before flare peak (panel b). At the time when the first brightening is visible in synthetic AIA emission (panel c, see also Figure \ref{fig9}) the rise speed has increased to about 40 km/s. The CME speed visible on the optical depth of $10^{-4}$ surface reaches $800$~km/s (panel e). Panel f) shows the RMS and 99.99$^{th}$ percentile velocity (mostly selecting the erupting MFR)) as function time, vertical dotted lines indicate the times shown in panels a) - e).}
    \label{fig12}
\end{figure*}

Figure \ref{fig12} shows the vertical velocity extracted on a corrugated optical depth $10^{-4}$ surface. While this optical depth usually corresponds to the lower chromosphere with a height of about 1~Mm above optical depth of unity, this surface is elevated by multiple Mm in the filament channel before the eruption and does trace cool material being ejected during the CME. A weak upflow velocity is visible in a snapshot $1200$~seconds before peak flare (panel a). While this velocity does not stand out in terms of amplitude, it does persist along the pre-eruption MFR and associated filament. This is the first indication of the start of a slow rise phase before the flare. At $400$~seconds before peak flare the upflow velocity starts standing out in amplitude by reaching an upflow speed of $10$~km/s (panel b), this corresponds to the time when there is a HFT detached from the photosphere present underneath the MFR (Figure \ref{fig10}f). At $155$~seconds before (point of onset) the flare this speed surpasses $40$~km/s and transitions quickly into more than $800$~km/s for the erupting CME (panels d and e). The slow-rise phase can be seen up to 1 hour before the flare, when looking at synthetic AIA $304$\AA\,emission in the side views (not shown here), but the corresponding velocities are too small to stand out in a display similar to Figure \ref{fig12}. 

\begin{figure*}
    \centering
    \resizebox{\hsize}{!}{\includegraphics{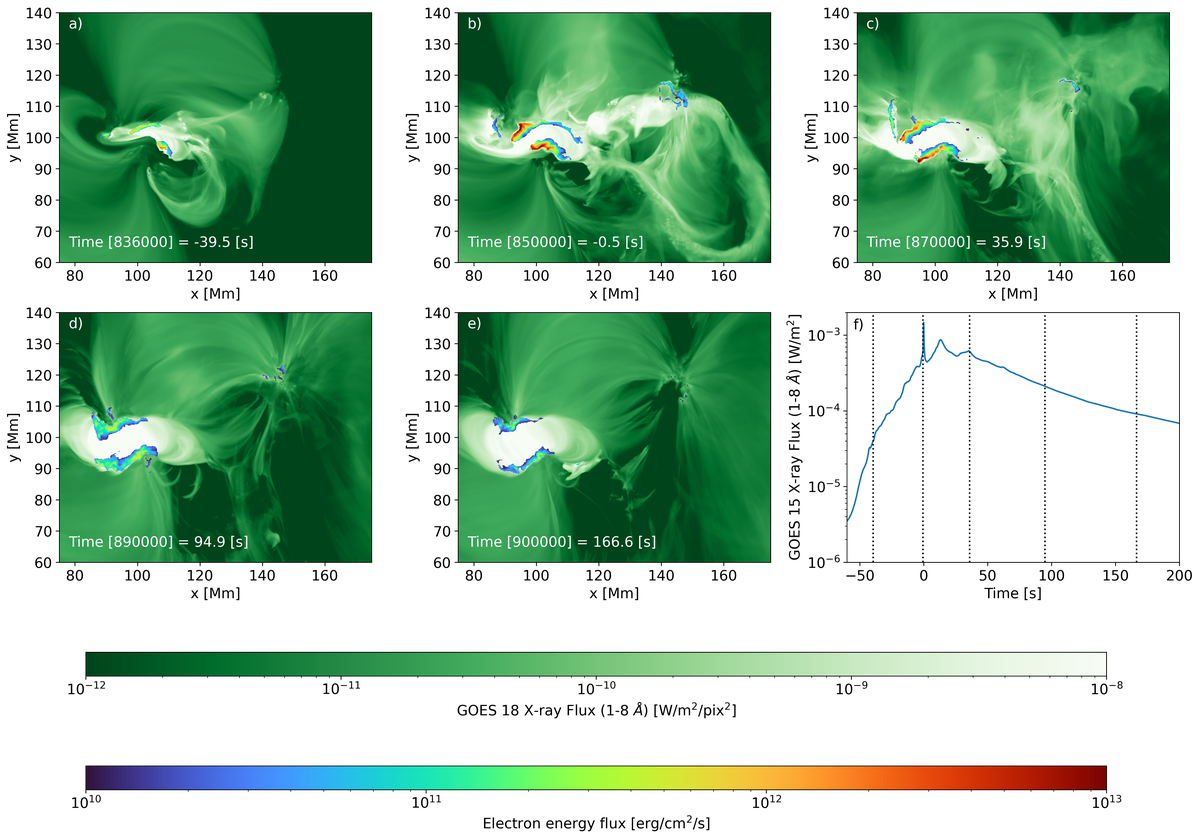}}
    \caption{ Time evolution of GOES $1-8$\AA\, emission and electron energy flux (free streaming limited Spitzer conduction) into the flare ribbons. Most of the energy is released over a time span of about 1.5 minutes, the electron energy flux reaches values of $10^{13}~\mbox{erg}~\mbox{cm}^{-2}~\mbox{s}^{-1}$.}    
    \label{fig13}
\end{figure*}

Figure \ref{fig13} shows the evolution of flare ribbons and synthetic GOES emission during the impulsive and gradual phase of the flare. Flare ribbons are indicated through the electron energy flux (parameterized through free-streaming limited Spitzer conduction) at a height of $1$~Mm above the photosphere. Overall this simulated flare does show mostly the topology of a two-ribbon flare with a pair of separating ribbons with hot, dense  X-ray emitting loops in-between, but at least early on (panel b) there is also a connection to the polarity P1. In this simulation the electron energy flux does reach values of up to $10^{13}~\mbox{erg}~\mbox{cm}^{-2}~\mbox{s}^{-1}$, which is a consequence of the relatively short duration of the flare as discussed in section \ref{sec:duration}.

\subsubsection{Photospheric signatures of flare}
   
\begin{figure*}
    \centering
    \resizebox{\hsize}{!}{\includegraphics{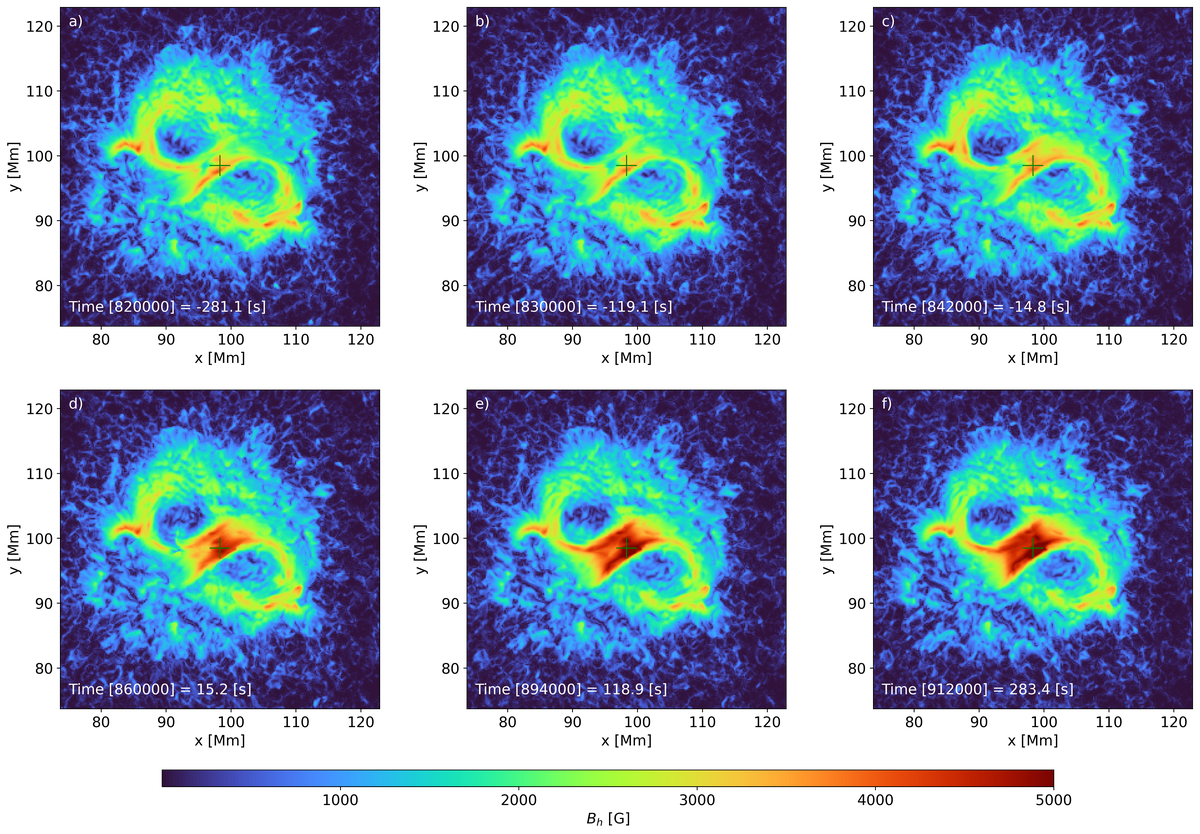}}
    \caption{Time evolution of the horizontal magnetic field strength at optical depth $\tau_{\rm 500}=0.1$. Panels a) - c) show the pre-flare values, panels d) -f) post flare values. There is step-function-like change of field strength in the region above the polarity inversion line, where the post-flare arcade is forming. The green '+' symbol indicated a location for with we show the time evolution of field strength in Figure \ref{fig15}.}
    \label{fig14}
\end{figure*}

\begin{figure*}
    \centering
    \resizebox{\hsize}{!}{\includegraphics{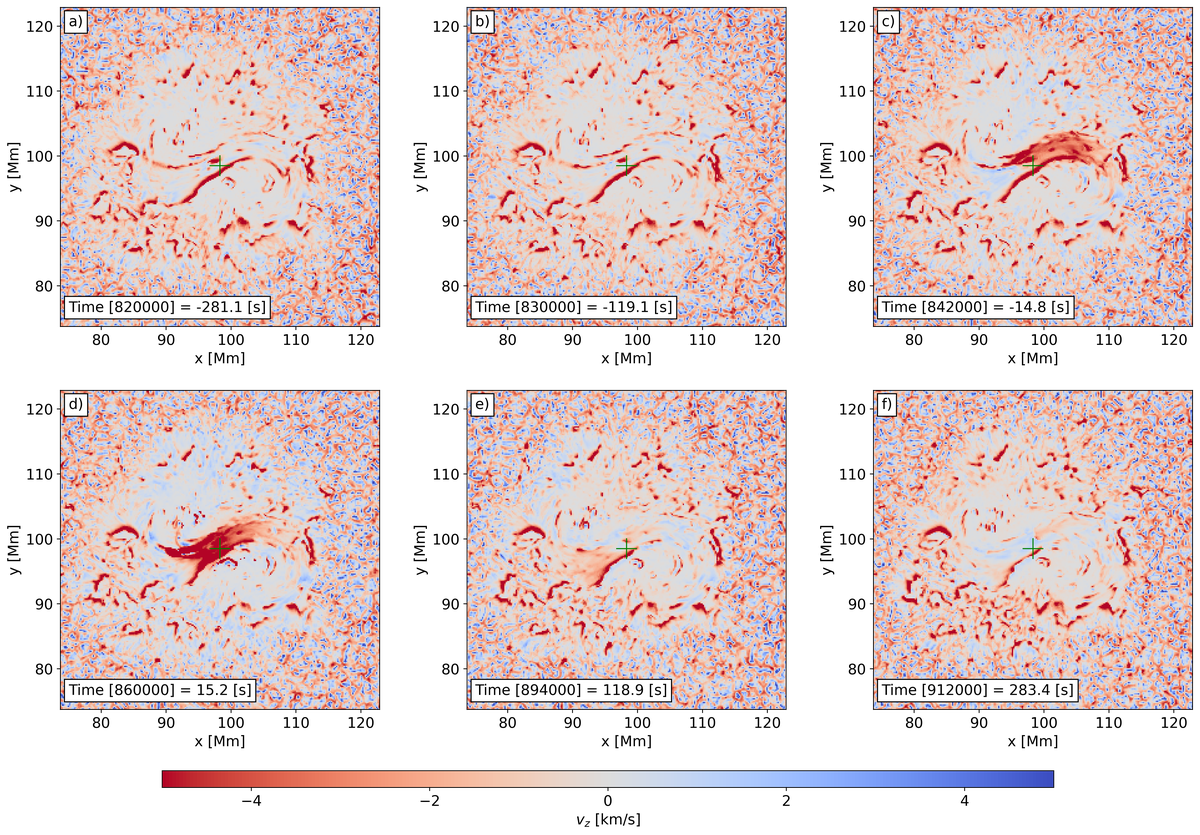}}
    \caption{Time evolution of the vertical velocity at optical depth $\tau_{\rm 500}=0.1$. Panels a) - c) show the pre-flare values, panels d) -f) post flare values. During the flare we find strong downflows above the polarity inversion line.}
    \label{fig15}
\end{figure*}

\begin{figure}
    \centering
    \resizebox{\hsize}{!}{\includegraphics{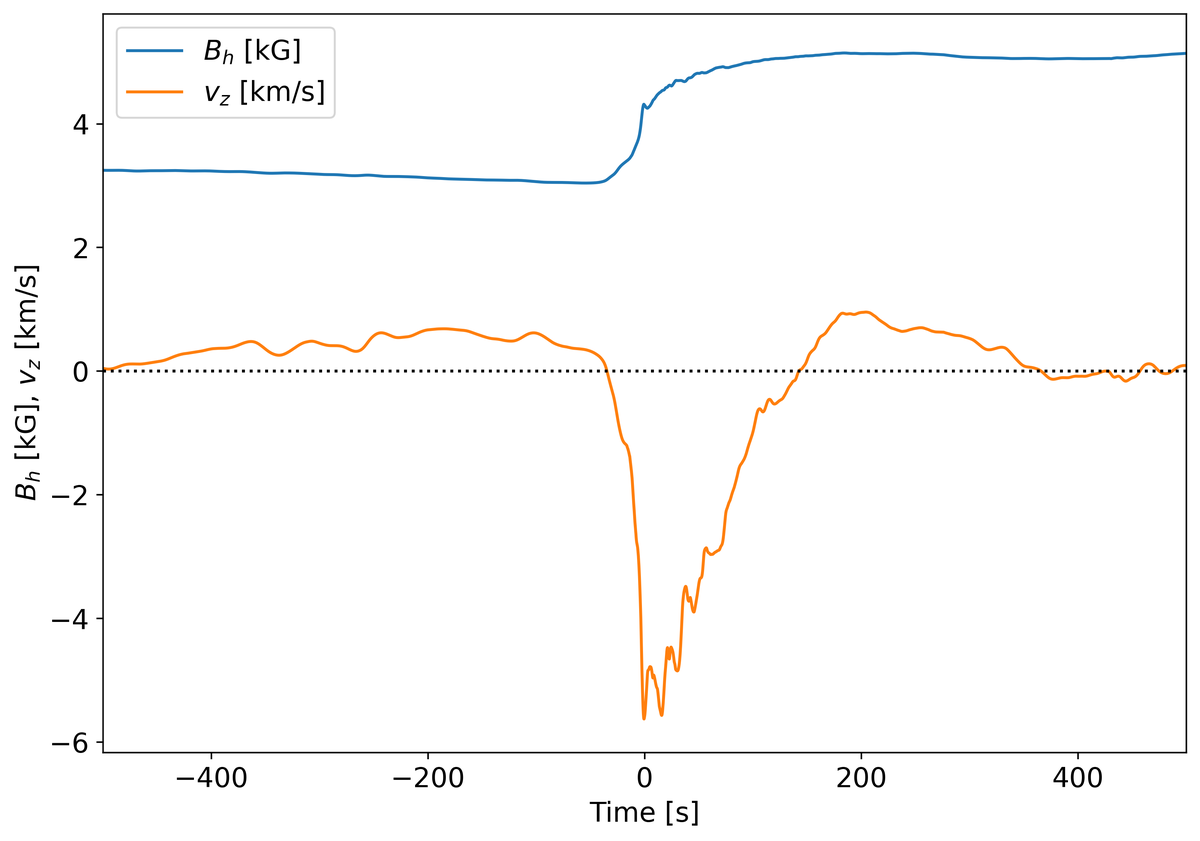}}
    \caption{Time evolution of horizontal field strength and vertical velocity for the position marked by a green '+' in Figures \ref{fig13} and \ref{fig14}.}
    \label{fig16}
\end{figure}

In this section we discuss flare induced changes visible in the photosphere. Figure \ref{fig14} shows the evolution of the horizontal magnetic field on the corrugated $\tau_{500}=0.1$ surface. While there is a mostly steady pre-flare (panels a-c) and post-flare (panels d-f) evolution, we do see a sharp change during the flare (panels c-d) in terms of an increase of the horizontal magnetic field near the PIL. This increase is accompanied with a strong downflow of about -5 km/s as shown in Figure \ref{fig15}. In Figure \ref{fig16} we show the time evolution of $B_h$ and $v_z$ for the position of the green cross shown in Figures \ref{fig14} and \ref{fig15}. The horizontal magnetic field shows a step function-like change similar to what has been reported for many observed flares \citet[e.g.][]{Wang:1992SoPh..140...85W,Sudol:2005ApJ...635..647S}. The magnitude of the persistent horizontal field change is about $1.5$~kG. This value is about 3 times larger than the largest observed longitudinal field change reported by \citet{Petrie:Sudol:2010:fieldchange}, which was observed for a limb flare, i.e. the longitudinal field did correspond to an actual horizontal magnetic field change. \citet{Wang:2012:AR11158_Bh} reported a change of the horizontal magnetic field from about $1300$~G to $1700$~G in AR 11158. The horizontal magnetic field present in our simulation is overall about a factor of $2-3$ stronger than the actual observed field, which may explain the overall stronger magnetic field change. The strong increase in horizontal field during the flare is consistent with the ''coronal implosion" scenario discussed by \citet{Hudson:2000:impl,Fisher:2012:backreaction}.

\subsubsection{Quasi-periodic Pulsations (QPPs)}
\label{sec:QPP}
\begin{figure}
    \centering
    \resizebox{\hsize}{!}{\includegraphics{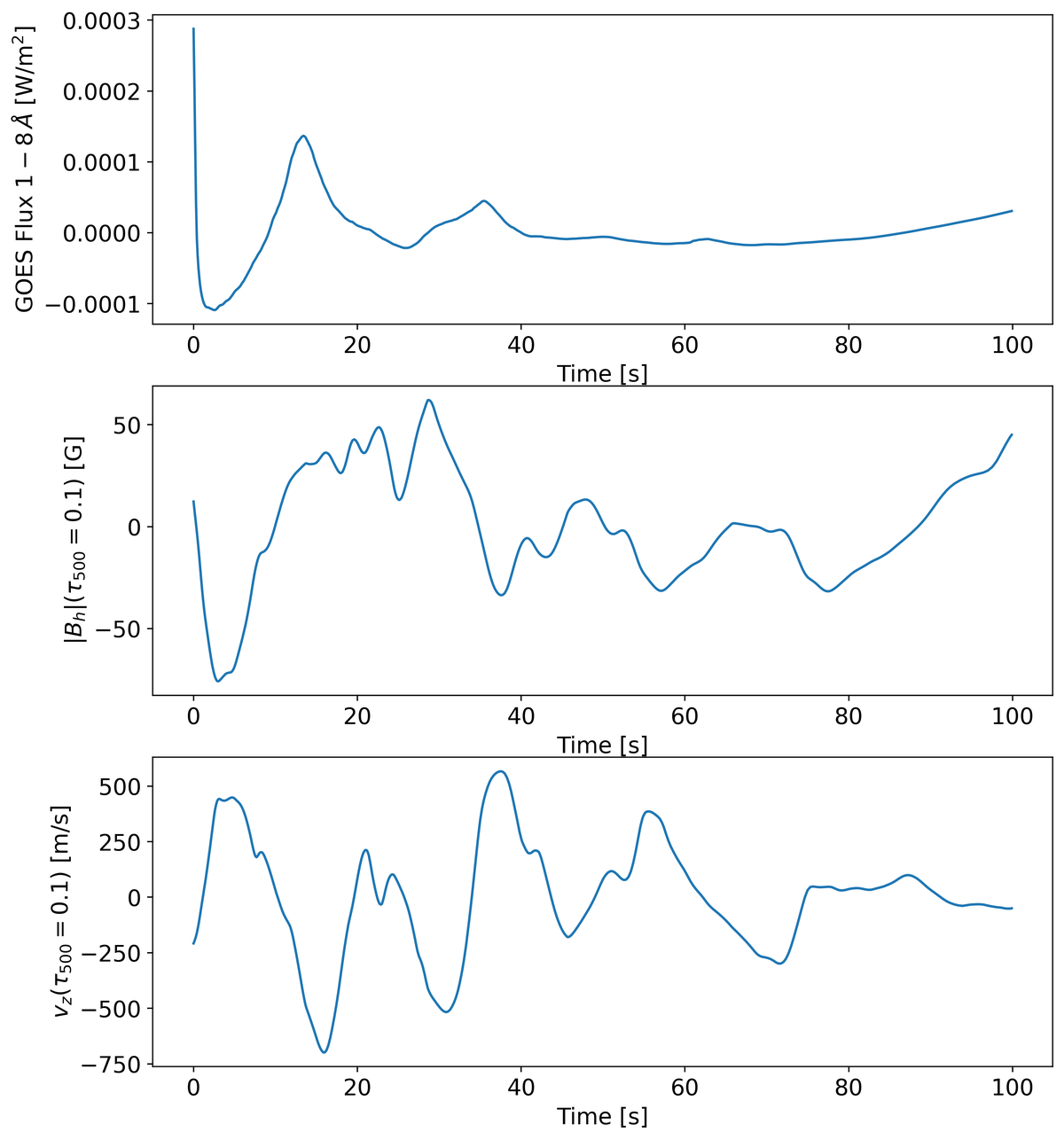}}
    \caption{Quasi periodic pulsations during the 100 seconds past the GOES peak: a) GOES X-ray flux, b) $\vert B_h\vert$ and c) $v_z$. The latter two correspond to the photospheric quantities shown in Figure \ref{fig16}. We subtracted a parabolic fit to remove the general trend.}
    \label{fig17}
\end{figure}

\begin{figure*}
    \centering
    \resizebox{\hsize}{!}{\includegraphics{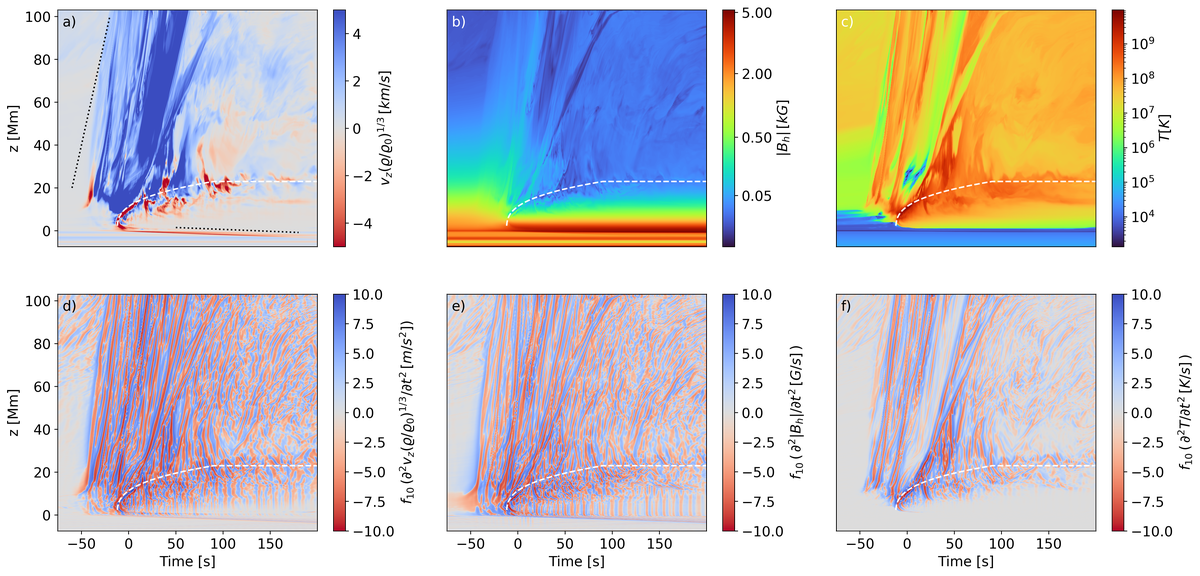}}
    \caption{Time-height slice of quasi periodic pulsations. Panels a) to c) show unfiltered quantities: a) Vertical velocity scaled by $(\varrho/\varrho_{\rm Phot})^{1/3}$ , b) $\vert B_h\vert$ and c) temperature. Panels d) to f) show the second time derivative of these quantities using a non-linear scaling as described in the text to highlight the oscillatory component of the signal. While the quasi periodic pulsations are present in the entire flaring corona, including the CME, their amplitude is strongest in the reconnection region indicated by a dashed white line. Standing oscillations with a longer duration are present in the post-flare loops in the lowermost 10~Mm of the corona. In the subphotospheric region we find a downward propagating pressure pulse. The dotted lines in panel a) indicate propagation speeds of $2,000$/$-15$~[km/s], respectively.}
    \label{fig18}.
\end{figure*}

\begin{figure}
    \centering
    \resizebox{\hsize}{!}{\includegraphics{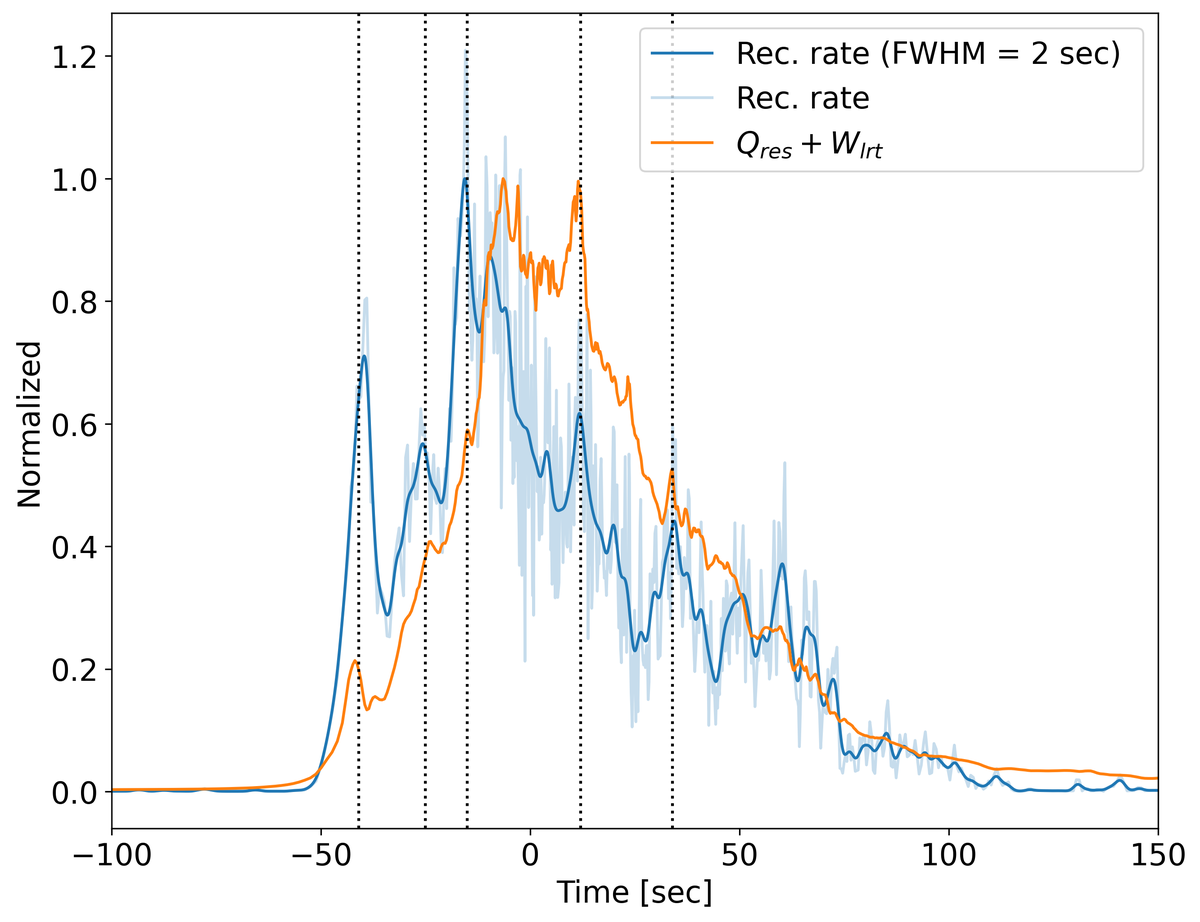}}
    \caption{Normalized profiles of flare ribbon derived reconnection rate (blue) and the total coronal energy release during the flare (sum of resistive heating and Lorentz force work) (orange). We indicated by dotted black lines times at which features in both quantities coincide. The solid blue line shows the reconnection rate smoothed with a Gaussian of 2 seconds FWHM, the faint blue line is the unfiltered reconnection rate.}
    \label{fig19}.
\end{figure}

Inspecting Figure \ref{fig16} carefully, we can identify a high frequency oscillation present in both magnetic field and vertical velocity during the impulsive phase of the flare. In Figure \ref{fig17} we enhance the oscillation signal during the 100 seconds following the GOES peak by subtracting the general trend through a parabolic fit. This is shown in panels b) and c). In panel a) we show the GOES X-ray flux filtered in a similar way. All three quantities do show oscillations in the 10-20 seconds range, the photospheric quantities show also shorter periods. The fact that these oscillations are present in both, photospheric quantities as well as X-ray emission that is an integral quantity, suggests that the underlying oscillation has to involve a larger volume in the corona. To demonstrate that we show in Figure \ref{fig18} time-height slices for the same horizontal position as shown in Figure \ref{fig17}. Since the velocity varies by several orders of magnitude from convection zone to corona, we show the quantity $v_z\,(\varrho/\varrho_{\rm Phot})^{1/3}$, where $\varrho_{\rm Phot}=3\cdot 10^{-7} {\rm g}\,{\rm cm}^{-3}$ is a typical photospheric density. Panels a) to c) show the unfiltered quantities. The vertical velocity (panel a) does show the outgoing CME and an upward propagating pattern of up- and downflows, which indicates the growing post flare arcade and upward propagation of the reconnection site (approximately indicated by the white dashed lines). In the convection zone we find a downward propagation of the flare induced photospheric downflow with the speed of sound. The horizontal magnetic field does show an enhancement in the lowermost $5$~Mm above the photosphere, while there is a reduction throughout the corona above $5$~Mm. Note that the horizontal stripes in the convection zone are due to magneto-convection features that evolve over time-scales much longer than shown here. The temperature in panel c) shows before the flare the cool filament channel reaching to a height of about $10$~Mm. The CME contains a mix of hot and cool plasma. Panels d) to f) show the second time derivatives of these quantities. In order to show a wider dynamical range we display a non-linear scaling using the function:
\begin{equation}
    f_s(x)={\arcsinh}\left({\sinh}(s)*\frac{x}{\max(x)}\right)
\end{equation}
with a parameter of $s=10$. We did filter out variations shorter than 2 seconds to keep the figure readable. We can identify oscillation that propagate outwards with the CME part of the simulation domain and reach towards the top of the simulation domain. Oscillations present underneath the post-flare arcade have more the appearance of a standing oscillation. Since a significant amount of the GOES X-ray emission comes from the hot post-flare loops (see Figure \ref{fig13}), it is possible that these standing loop oscillations can modulate the GOES emission. The reconnection rate shown in Figure \ref{fig8} does also show variability over a wide range of time-scales. We filtered out variability on scales shorter than 2 seconds similar to Figure \ref{fig18} for better visibility. Also here we do find quasi-periodic pulses in the 10-20 seconds range. Since the reconnection rate is derived from the flare ribbon progression, which could be modulated by oscillations in the post-flare arcade, we also do see variability that is clearly linked to intrinsic variation in the coronal reconnection rate. Figure \ref{fig19} shows an overlay of normalized profiles for (flare ribbon derived) reconnection rate and coronal energy release (sum of numerical resistive heating and Lorentz force work). Black dotted lines indicate times at which features are visible in both quantities (suggesting a coronal origin), but there is also variability in both quantities without a clear correspondence. The faint blue line shows the unfiltered reconnection rate with sub-second variability.  

The downward propagating pulse in the sub-photospheric region is possibly linked to sunquakes \citep{Donea:1999:sunquake,Kosovichev:2006:sunquake}, but we did not perform any analysis to further investigate that connection. As we show in Figure \ref{fig16}, the momentum transfer into the photosphere during the magnetic implosion is a possible scenario \citep{Fisher:2012:backreaction}.

\section{Conclusions}\label{sec:concl}
We presented a numerical setup inspired by the active region AR 11158. While our approach does not account flux emergence, it captures the collisional shearing present during the evolution of the active regions. This was accomplished by imposing sunspot motions that capture the evolution of the observed sunspot centroid positions. The main findings from this investigation are summarized below.

\subsection{Flare strength and efficiency}
The collisional shearing alone present in AR 11158 is sufficient to build up the free energy needed for an X-flare. Our simulation reached more than $4\cdot 10^{32}$~erg of free energy out of which about $50\%$ were released an a single X-flare. The continuing shearing caused overall a series for more than 10 flares releasing about $3\cdot 10^{32}$~erg. Flares driven by collisional shearing can be very efficient. Similar to our earlier works \citep{Rempel:etal:2023} we found that up to $50\%$ of the stored free energy can be released by this process. This high efficiency is only found for the first flare in the series, all following flares release smaller amounts of energy ($5-20\%$ of the stored free energy). Collisional shearing alone can lead to a series of multiple flares.
 
\subsection{Pre-flare MFR and trigger}
The evolution of the cPIL involves both cancellation and re-emergence of magnetic flux confirming what was discussed in the work of \citet{Chintzoglou:etal:2019}. When the central polarities begin to collide, flux cancellation does not lead to the formation of a MFR due to the lack of shear. We note that during the early stage of collision the coronal magnetic field is still close to potential due to the initial setup of this simulation. The MFR buildup happens during a second cancellation epoch starting about 2 hours before the X-flare when the system is already heavily sheared. The MFR is slowly rising during the buildup phase until it reaches the height where the decay index exceeds a value around $1.5$, suggesting the torus instability as flare trigger. We do find a slow rise phase preceding the flare, best visible in the vertical velocity of plasma in the filament channel. Early indications can be seen about $20$ minutes before the flare, a clear velocity above the typical background is present $5$ minutes before the flare. Looking at side views of synthetic AIA $304$\AA\, emission allows to trace back the slow-rise phase to about one hour before the flare. The slow rise phase coincides with the time of buildup of twist in the pre-flare MFR, suggesting that the rise phase could be an indicator of MFR formation and approach to the threshold of torus instability, similar to the findings of \citet{Cheng:etal:2023:slow-rise,Xing:etal:2024:slow-rise}. 
 
\subsection{Flare duration}
The duration of the simulated flares is in the range of 1.5 to 10 minutes, which falls into the lower end of the observed solar flare durations, suggesting an up to an order of magnitude too high reconnection rate in current simulations. This impacts other flare properties such energy fluxes into flare ribbons, the propagation speed of flare ribbons and CME speeds. Reconnection in the simulation is due to numerical resistivity, which does lead (due to nonlinearities) to fast reconnection. The resolution of the current sheet is insufficient to allow for fragmentation, leading to essentially single X-line reconnection with a rate of $R\sim 0.1$. While there are certainly solar flares that happen this fast, the typical flares appears to be about an order of magnitude slower. Such behavior is expected in the coronal parameter regime, where multiple X-line reconnection can lead to on average slow reconnection with a rate of $R\sim 0.01$ \citep{Uzdensky:2010:ReconnectionPlasmoid,Huang:Bhattacharjee:2010:HighLundquist,Ji:etal:2022}. As we showed in Figures \ref{fig3} and \ref{fig9}, most of the free energy is stored within 5~Mm above the photosphere and the flare reconnection starts at a height of just 2~Mm above the photosphere. These regions do have a high Alfv\'en velocity (several $10,000$ km s$^{-1}$) and consequently can lead to high reconnection rates. It is also feasible that slower flares happen in configurations where the free energy is stored higher up in the corona, where the Alfv\'en velocity is lower. 

\subsection{Flare imprint on transition region and lower atmosphere}
The X flare we analyzed has primarily the morphology of a two ribbon flare, with initially some connection to the polarity P2. The propagation of the flare ribbons and the derived reconnection rate are about a factor of $3-10$ higher than for a typical solar flare of comparable strength \citep{Tamburri:2024}, which is in line with the too short flare duration. Similarly electron energy fluxes into the flare ribbons are with $10^{13}$ erg cm$^{-2}$ s$^{-1}$ too high for typical solar flares. Estimates of non-thermal energy fluxes for strong solar flares are typically in the $10^{10}-10^{11}$ erg cm$^{-2}$ s$^{-1}$ range \citep{Kuridze:2015ApJ...813..125K,Kleint:2016ApJ...816...88K,Kowalski:etal:2017,Kerr:2024ApJ...970...21K}, although values as high as $10^{13}$ erg cm$^{-2}$ s$^{-1}$ have been inferred for flares on M dwarfs \citep{Kowalski:2015:MDwarfFlare}.

We find a step-function like increase of the horizontal field strength in the photosphere associated with strong downflows in between the flare ribbons caused by the collapsing post-reconnection loops. The stepwise change of horizontal fields is qualitatively similar to observations \citep{Wang:1992SoPh..140...85W,Sudol:2005ApJ...635..647S}, although we do find associated field strength more than a factor of 2 larger than the strongest field changes observed \citep{Petrie:Sudol:2010:fieldchange}. It is also possible that observed field strength may be underestimated by a similar factor \citep{Petrie:2025ApJS..278...55P} due to calibration uncertainties of HMI magnetograms.

\subsection{Quasi-periodic pulsations (QPPs)}
During the flare we find oscillatory behavior throughout the corona reminiscent of QPPs \citep[see, e.g.][for a recent review]{QPP:2021SSRv..217...66Z} and a pulse propagating into the convection zone, which could be related to sunquakes \citep{Donea:1999:sunquake,Kosovichev:2006:sunquake}. The QPPs do involve a multitude of frequencies, we find variations on sub-second scales in the reconnection outflows to durations in the 10-20 seconds range. QPPs are visible throughout most of the flaring corona and can be consequently seen in many different observables (photospheric magnetic field and velocity, GOES emission, reconnection rate derived from flare ribbon propagation). We find evidence for standing oscillations in the post flare loop system as well as a connection to bursts in the coronal reconnection rate. Recent work by \citet{Ashfield:2026Nat:QPP} demonstrated oscillatory reconnection as driver for QPPs observed in HXR/chromospheric data. While they also found MHD loop oscillations, they could exclude them as process behind the observed HXR/chromospheric QPPs. Since our current flare model does not contain particle beams we cannot address these aspects of QPPs. However, both their work and our simulation indicates that QPPs are likely the consequence of multiple processes in a strongly coupled system and different observables may be impacted by different drivers. \hspace{1cm}\\

Overall this study shows that collisional shearing is a potent process to power major solar eruptions. While the MHD treatment presented here can capture many aspects of flares from photosphere to corona, there are also clear limitations of this treatment. We found that reconnection rates are systematically too high by about an order of magnitude compared to the typical solar flare. Furthermore, a more consistent description of flares will require an accounting for non-thermal energy and non-EUV radiative losses through deposition of energy beneath the transition region. Currently efforts are underway that improve the treatment of the transition region and allow to account for non-thermal energy transport and deposition through a parameterized treatment.
  
\begin{acknowledgements}
This material is based upon work supported by the NSF National Center for Atmospheric Research, which is a major facility sponsored by the U.S. National Science Foundation under Cooperative Agreement No. 1852977. We would like to acknowledge high-performance computing support from the Derecho system (doi:10.5065/qx9a-pg09) provided by the NSF National Center for Atmospheric Research (NCAR), sponsored by the National Science Foundation. VAPOR is a product of the National Center for Atmospheric Research’s Computational and Information Systems Lab. Support for VAPOR is provided by the U.S. National Science Foundation (grants 03-25934 and 09-06379, ACI-14-40412), and by the Korea Institute of Science and Technology Information. M.R. is partially supported through the NASA Multi-slit Solar Explorer (MUSE) mission through the contract 80GSFC21C0011 and thanks the MUSE science team for many inspiring discussions from which this work has benefited. G.C. acknowledges support from NASA Heliophysics Guest Investigator (HGI) grant 80NSSC25K7691. We thank Amy Winebarger and Biswajit Mondal for providing updated GOES 18 response functions. We thank Yuhong Fan for comments on the manuscript. This research was supported by the International Space Science Institute (ISSI) in Bern, through ISSI International Team project 'Understanding the Onset of Solar Eruptions' (ISSI Team project \#24-606).
\end{acknowledgements}

\bibliography{papref_m}
\bibliographystyle{aasjournal}

\end{document}